\documentclass[nohyper,12pt,letterpaper]{JHEP3}

\usepackage{amsmath,amsfonts,amssymb,xspace,multirow}

\DeclareMathOperator{\im}{Im}
\DeclareMathOperator{\re}{Re}
\DeclareMathOperator{\Ad}{Ad}

\newcommand{\che}{}

\newcommand{\cbra}{_{(t, \bar{t})}\!\langle x,\lambda |}
\newcommand{\bcohb}{\langle x^I |}

\newcommand{\psiv}{\Psi_{\mathrm V}}
\newcommand{\psiw}{\Psi_{\mathrm W}}

\newcommand{\I}{{\mathrm i}}

\newcommand{\ctM}{{\widetilde{\mathcal M}}}

\newcommand{\chM}{{\wh{\mathcal M}}}
\newcommand{\cM}{{\mathcal M}}

\newcommand{\cL}{{\mathcal L}}
\newcommand{\cK}{{\mathcal K}}
\newcommand{\cT}{{\mathcal T}}

\newcommand{\cH}{{\mathcal H}}

\newcommand{\cO}{{\mathcal O}}
\newcommand{\cJ}{{\mathcal J}}
\newcommand{\cV}{{\mathcal V}}

\newcommand{\tP}{{\widetilde{P}}}
\newcommand{\tQ}{{\widetilde{Q}}}
\newcommand{\tG}{{\widetilde{G}}}

\newcommand{\tK}{{\widetilde{K}}}

\newcommand{\ahol}{{\mathrm{ah}}}
\newcommand{\hol}{{\mathrm{hol}}}
\newcommand{\BCOV}{{\mathrm{BCOV}}}

\newcommand{\hp}{{\wh{p}}}
\newcommand{\hq}{{\wh{q}}}

\newcommand{\bx}{{\bar{x}}}

\newcommand{\bt}{{\bar{t}}}
\newcommand{\bi}{{\bar{i}}}
\newcommand{\bj}{{\bar{j}}}

\newcommand{\bcL}{{\bar{\cL}}}
\newcommand{\bcT}{{\bar{\cT}}}

\newcommand{\eg}{\ti{e.g.}\xspace}

\newcommand{\R}{\ensuremath{\mathbb R}}
\newcommand{\C}{\ensuremath{\mathbb C}}
\newcommand{\PP}{\ensuremath{\mathbb P}}
\newcommand{\Z}{\ensuremath{\mathbb Z}}

\newcommand{\ihalf}{\ensuremath{{\frac{\I}{2}}}}
\newcommand{\ieighths}{\ensuremath{{\frac{\I}{8}}}}
\newcommand{\half}{\ensuremath{{\frac{1}{2}}}}
\newcommand{\sixth}{\ensuremath{{\frac{1}{6}}}}
\newcommand{\qtr}{\ensuremath{{\frac{1}{4}}}}

\newcommand{\N}{{\mathcal N}}

\newcommand{\kahler}{{K\"ahler}\xspace}

\newcommand{\abs}[1]{\lvert#1\rvert}
\newcommand{\norm}[1]{\lVert#1\rVert}
\newcommand{\IP}[1]{\langle#1\rangle}

\newcommand{\wh}[1]{\widehat{#1}}
\newcommand{\dwrt}[1]{\frac{\partial}{\partial#1}}

\newcommand{\tfg}{\widetilde{\fg}}
\newcommand{\fg}{{\mathfrak g}}

\newcommand{\ket}[1]{{\vert #1 \rangle}}
\newcommand{\bra}[1]{{\langle #1 \vert}}

\newcommand{\ti}[1]{\textit{#1}}

\newcommand{\tm}{Teichm\"uller\xspace}
\newcommand{\sw}{Schr\"odinger-Weil\xspace}

\newcommand{\pa}{\partial}

\newcommand{\IR}{\mathbb{R}}

\def\bea{\begin{eqnarray}}
\def\eea{\end{eqnarray}}
\def\be{\begin{equation}}
\def\ee{\end{equation}}
\def\ba{\begin{align}}
\def\ea{\end{align}}
\def\bse{\begin{subequations}}
\def\ese{\end{subequations}}
\def\1F1{{}_1\!F_1}
\def\2F0{{}_2\!F_0}

\preprint{hep-th/0607200\\LPTENS-06-26}

\title{Topological wave functions and heat equations}

\author{Murat G\"unaydin,
\footnote{Email: {\tt murat@phys.psu.edu}, {\tt neitzke@post.harvard.edu},
{\tt pioline@lpthe.jussieu.fr}} \\
Department of Physics, Penn State University, \\
University Park, PA 16802, USA}

\author{Andrew Neitzke,$^*$\\
School of Natural Sciences, Institute for Advanced Study, \\
Princeton, NJ 08540, USA}

\author{Boris Pioline $^*$\\
LPTHE, Universit{\'e}s Paris VI \& VII, 4 place Jussieu, F-75252 Paris\\
{\it and} \\
LPTENS, D{\'e}partement de Physique de l'ENS, 24, rue Lhomond,
F-75005  Paris}

\abstract{
It is generally known that the holomorphic anomaly equations
in topological string theory reflect the quantum mechanical nature
of the topological string partition function.  We present two
new results which make this assertion more precise:
(i) we give a new,
purely holomorphic version of the holomorphic anomaly equations,
clarifying their relation to the heat equation satisfied by
the Jacobi theta series;
(ii)
in cases where the moduli space is a Hermitian symmetric tube domain
$G/K$, we show that the general solution of the anomaly equations
is a matrix element
$\IP{\Psi \vert g \vert \Omega}$ of the Schr\"odinger-Weil representation of a Heisenberg extension of $G$,
between an arbitrary state $\bra{\Psi}$ and a particular
vacuum state $\ket{\Omega}$.
Based on these results, we speculate on the existence of
a one-parameter generalization of the usual topological amplitude,
which in symmetric cases transforms in the smallest unitary representation
of the duality group $G'$ in three dimensions, and
on its relations to hypermultiplet couplings,
nonabelian Donaldson-Thomas theory and black hole degeneracies.}

\begin{document}

\renewcommand{\baselinestretch}{1.25}
\small\normalsize

\section{Introduction}

On top of its rich mathematical structure, topological string theory
offers the most practical way of computing certain higher-derivative
corrections to the low energy effective action in $\N=2$ compactifications
of physical string
theory.  It has played a central role in recent studies of subleading
corrections to the Bekenstein-Hawking entropy of BPS black
holes \cite{LopesCardoso:1998wt,
LopesCardoso:1999cv,LopesCardoso:1999ur}.  A particularly striking development
is a direct relation between the degeneracies of black hole
micro-states and the topological string partition function,
conjectured by Ooguri-Strominger-Vafa (OSV) \cite{Ooguri:2004zv}.

The topological string
partition function satisfies ``holomorphic anomaly'' equations
derived by Bershadsky-Cecotti-Ooguri-Vafa (BCOV) \cite{Bershadsky:1993ta,Bershadsky:1994cx},
which describe its dependence on a ``background'' point of moduli space.
These equations are best understood by viewing the topological string
partition function as the wave function of a background-independent state
in a certain Hilbert space, expressed in a background-dependent basis
of coherent states \cite{Witten:1993ed}.  In particular, as
emphasized in \cite{Verlinde:2004ck}, for the application to the OSV
conjecture one should choose a basis of coherent states
corresponding to a ``real polarization'' of the wave function.
The purpose of the present work is to tighten the wave function
interpretation of the holomorphic anomaly equations, by recasting
these equations in a purely holomorphic way, and representing their
solution as a matrix element in an appropriate representation
of the continuous global symmetry group of the moduli space, when such a symmetry exists.

To motivate our approach, recall that the wave function interpretation
of the holomorphic anomaly equations was originally suggested \cite{Witten:1993ed} by analogy with the
heat equations
\begin{align}
\che \left[\I \dwrt{t} - \frac{\partial^2}{\partial y^2} \right] \theta(t,y) &= 0, \label{heatth} \\
\che \dwrt{\bar t}\  \theta(t,y) &= 0, \label{heatth0}
\end{align}
obeyed by the classical Jacobi theta series $\theta (t,y)$.
However, the anomaly equations as written in \cite{Bershadsky:1994cx} are
not quite identical to \eqref{heatth}, \eqref{heatth0}.  In particular,
the topological string
partition function $\Psi_{\rm BCOV}$ is neither purely holomorphic nor
purely antiholomorphic.
So the precise analogy between $\Psi_{\rm BCOV}$ and $\theta(t,y)$
has remained slightly obscure.

In Section \ref{sec-revisited} of this paper, we draw a precise analogy
between the topological string partition function $\Psi_{\rm BCOV}$
(or rather, its variant $\psiv(t,\bar t,x)$ as defined in
\cite{Verlinde:2004ck}), and the complex conjugate of the
``canonical'' Jacobi theta function $\Theta(t,\bar t, \bar x)$,
\be
 \label{analogy}
\psiv\left(t,\bar t,x\right) \sim \Theta^*(t,\bar t, x),
\ee
where $\Theta(t,\bar t,\bar x)$ is a non-holomorphic modular function which
differs from the classical holomorphic $\theta(t,y)$ by a simple
transformation,
\begin{align}
 \label{analogy2-1}
\Theta(t,\bar t,\bar x) &= \sqrt{\im t}\ \exp\left(\frac{y^2}{\im t}\right)\
\theta(t, y), \\
y &= (\im t) \bar x. \label{analogy2-2}
\end{align}
Namely, by transformations of $\psiv^*$ similar to \eqref{analogy2-1}, \eqref{analogy2-2},
described in Section \ref{sec-holomorphic},
we produce a modified partition function
$\Psi_\hol$, which is purely holomorphic and obeys a
heat equation \eqref{holom-a-2} closely analogous to \eqref{heatth}.
These transformations, valid in complete generality, provide a more concrete
footing for the widely accepted idea that the anomaly equations
of the topological string partition function should be analogous
to the heat equation of a theta function.
Along the way we also clarify a few details which have not appeared
explicitly in the literature, in particular the relation between
the one-loop holomorphic anomaly and the metaplectic correction to geometric quantization.

Importantly, our transformation from $\psiv^*$ to $\Psi_\hol$
introduces a dependence on a choice of symplectic basis, which is not invariant under large biholomorphisms of the
Calabi-Yau threefold; consequently, the transformation of the holomorphic partition function $\Psi_\hol$
under large biholomorphisms is more complicated than that of the original $\psiv$, as we discuss
in Section \ref{sec-modular-general}.
This parallels the fact that the holomorphic $\theta(t,y)$ has a more
complicated transformation under $SL(2,\Z)$ than does
the non-holomorphic $\exp \left(\frac{y^2}{\im t}\right) \theta(t,y)$.

In Section \ref{sec-symmetric}, we specialize the discussion to cases where
the moduli space is a symmetric special K\"ahler manifold $G/K$.
While such cases are very special (in particular they require the absence
of genus 0 worldsheet
instanton corrections), they allow us to develop a second aspect of the analogy
between $\Psi$ and $\theta(t,z)$, representing the topological amplitude
as a matrix element in a unitary representation of an extension of $G$.  To appreciate
this point, recall the standard interpretation  of $\theta(t,z)$ in
representation theory \cite{MR781735,MR1634977}:
One considers the semi-direct product $\tG = SL(2,\R)\ltimes H$,
where $H$ is the three-dimensional
Heisenberg algebra $[\wh{q},\wh{p}] = Z$, and $G = SL(2,\R)$ acts on $H/Z$
in the fundamental representation.  Both $G$ and $H$ act naturally on the Hilbert space obtained by
quantizing $\R^2$, which we call the ``\sw representation'' of $\tG$.
Then $\theta(t,z)$ may be written as a matrix element in this representation,
\be
\label{matth}
\theta(t,y) = \langle \Psi |
e^{y \wh{p}} e^{t T} | \Omega \rangle,
\ee
where $T$ is the positive root generator $\bigl( \begin{smallmatrix} 0 & 1 \\ 0 & 0 \end{smallmatrix} \bigr)$ of $\mathfrak{sl}(2,\IR)$, $\ket{\Omega}$
is a state annihilated by $\wh{q}$, and $\bra{\Psi}$ is a state invariant under
a subgroup of $SL(2,\Z)$.
In this interpretation the heat equation \eqref{heatth} becomes
a consequence of an operator identity in the \sw representation,
\be
\label{sweq}
ZT - {\wh{p}}^2 = 0,
\ee
if we choose the central generator to act as $Z=\I$.  Note that the
heat equation would hold no matter which state $\ket{\Psi}$ we choose;
the most general solution is obtained
by choosing an arbitrary $\ket{\Psi}$.

We show that a similar story holds for the topological string partition function,
as one would expect from its wave function interpretation:
there is a 1-1 correspondence between states $\ket{\Psi}$ in the \sw representation of an appropriate group
$\tG = G\ltimes H$ and solutions to the anomaly
equation \eqref{holom-a-2}.
Given $\ket{\Psi}$, the corresponding solution can
be written as a matrix element similar to \eqref{matth}, given in \eqref{psi-gt}.
From this point of view, the holomorphic anomaly equations
are consequences of identities \eqref{op-anom-0} in the \sw representation,
analogous to \eqref{sweq} (but slightly more complicated.)

In Section \ref{sec-3d} we discuss a third aspect of the analogy between $\Psi$ and $\theta$,
which may be motivated as follows.
Equation \eqref{sweq} can be obtained by
noting that the non-reductive group $SL(2,\R) \ltimes H$ is naturally obtained as a subgroup
of the more familiar $Sp(4,\R)$, and the \sw representation of $SL(2,\R) \ltimes H$
is contained in the metaplectic (Weil)
representation of $Sp(4,\R)$.  From this point of view, \eqref{sweq} is just one of a large
class of quadratic operator equations which hold in the Weil representation of $Sp(4,\R)$.
We note that a very similar situation obtains for the topological string:
the identities \eqref{op-anom-0} involving generators of $\tG$ come
directly from relations in
the ``minimal representation'' of a larger and more familiar group $G'$.
We also briefly discuss
the automorphic version of this relation, which in the $Sp(4,\R)$ case gives
rise to the notion of Fourier-Jacobi
coefficients.

We emphasize that none of our rigorous results bear directly on the problem of \ti{computing}
the topological string partition function; they are, rather, a clarification of what kind of object
the partition function is.  However, they do tie in naturally to
some other directions of inquiry which were really the original motivation,
and which we discuss in Section \ref{sec-discussion}.

In particular, we note that the group $G'$ mentioned above arises
physically as the local isometry
group of the moduli space upon compactification from $d=4$ to $d=3$.
The group  $G'$ has been proposed as a spectrum generating symmetry
group  in $d=4$ supergravity, and its minimal unitary representation
studied in  connection with  black hole degeneracies (see
\cite{Gunaydin:2005mx} and references therein.)  For the relation to the
topological string, the most relevant point is that the minimal representation
of $G'$ contains a one-parameter family of \sw representations of $\tG$ with
varying values for the center $Z=\I \hbar$, with the representation
relevant for the topological string arising at $\hbar=1$.  This suggests the
existence of a ``generalized topological amplitude'', a one-parameter extension
of the usual perturbative topological amplitude, which (in the special cases
we consider) would be described by a state in the minimal representation
of $G'$.  We conjecture that this
generalized topological amplitude could control corrections to hypermultiplet
couplings
and be related to the nonabelian Donaldson-Thomas theory.
We also briefly discuss the appealing possibility
that the topological amplitude could literally be a theta function
(in the appropriate sense), which
would in principle determine the whole
nonperturbative partition function up to at most a finite-dimensional ambiguity.

For completeness, in Appendices A and B we discuss
a relation between the real-polarized wave function
and the ``holomorphic ambiguities'' introduced by BCOV in solving
the anomaly equations, and the form of the wave function in
a positive-definite holomorphic polarization.

Holomorphic rewritings of the anomaly equations in some special cases have appeared before
in the literature, \ti{e.g.} \cite{Hosono:1999qc}.  The anomaly equations and their quantum-mechanical
interpretation have also been reexamined in \cite{todorov-gq,Gerasimov:2004yx}.
Modular properties of the anomaly equations have recently been discussed in \cite{Huang:2006si},
and treated more generally in
\cite{Aganagic:2006wq}, which appeared while this paper was being finalized.  Parts of \cite{Aganagic:2006wq} overlap
with parts of Section \ref{sec-revisited} of this paper.

\section{The anomaly equations revisited} \label{sec-revisited}

In this section, we discuss various forms of the holomorphic anomaly equations.
Most of these have already appeared in the literature; our main purpose
is to introduce a new version of the topological string partition function
which is purely antiholomorphic, and whose dependence on the antiholomorphic
variables is governed by a heat-type equation given in \eqref{holom-a-2}
below.  Along the way, we also clarify some aspects of the more standard
presentations of the topological amplitude.  Our discussion is couched in
the language of the B model topological string on a Calabi-Yau threefold $X$,
but with suitable changes of notation and wording
it would apply equally well to the A model.

\subsection{BCOV's holomorphic anomaly equations}

Let $\cM$ denote the Teichm\"uller space of complex structures on $X$, equipped
with its Weil-Petersson metric $g$.  Let $\cT$ be the holomorphic tangent bundle to $\cM$, and $\cL$
the line bundle $H^{3,0}(X,\C)$ over $\cM$.

In terms
of local complex coordinates on $\cM$, a point
$t \in \cM$ may be written $(t^i,\bar t^{\bar i})$, with $i = 1, \dots, n_v$,
where $n_v = h^{2,1}(X)$.  We also work in a local trivialization of the line bundle
$\cL$, given by some local holomorphic section $\Omega(t)$.  Then let $K$ denote the K\"ahler potential,
defined by
\begin{equation}
\che K(t,\bar t) = - \log \norm{\Omega}^2 = - \log \I \int_X \Omega \wedge \overline{\Omega}.
\end{equation}
The metric compatible connection in $\cL$ is
\begin{equation}
\che D_i s = \left(\pa_i - \pa_i K\right) s.
\end{equation}
Moreover, $K$ also serves as a local \kahler potential for the metric on $\cM$:
\begin{equation}
g_{i \bar{j}} = \pa_i \bar{\pa}_\bj K.
\end{equation}

Following \cite{Bershadsky:1994cx}, we consider the generating function of
the genus $g$ correlation functions
$C^{(g)}_{i_1 \cdots i_n}$ of $n$ chiral fields, in the B model topological string theory on $X$:
\be
\label{BCOVW} \che
W(t^i,\bar t^{\bar i}; x^i,\lambda) = \sum_{g=0}^{\infty}
\sum_{n=0}^{\infty} \frac{1}{n!} \lambda^{2g-2}
\ C^{(g)}_{i_1 \cdots i_n}(t,\bar t)\  x^{i_1}
\cdots x^{i_n} + \left( \frac{\chi}{24}-1 \right)\log\lambda.
\ee
This is a formal series which need not converge; throughout this paper we will ignore this issue.
We consider $x^i$ as the coordinates of a vector $x^i \dwrt{t^i} \in \cT$.  Similarly we consider the
inverse topological string coupling $\lambda^{-1}$ as the coordinate of an underlying vector
$\lambda^{-1} \Omega \in \cL$,
and $C^{(g)}_{i_1 \cdots i_n}$ as the coordinate expression of an element in $(\cT^*)^n \otimes \cL^{2g-2}$.

The correlation functions $C^{(g)}$ are given by $C^{(g)}_{i_1\dots i_n}=0$ for $2g-2+n\leq 0$
(so $W$ does {\it not} incorporate the genus 0 and 1 vacuum amplitudes)
and
\be \label{chiral-C}
\che C^{(g)}_{i_1\cdots i_n}(t,\bar t) = \left\{ \begin{array}{ccc}
D_{i_1}\cdots D_{i_{n-3}} C_{i_{n-2} i_{n-1} i_n} & \mbox{for} & g=0, n \ge 3, \\
D_{i_1}\cdots D_{i_n} F_1 & \mbox{for} & g=1, n \ge 1, \\
D_{i_1}\cdots D_{i_n} F_g & \mbox{for} & g \ge 2.
\end{array} \right.
\ee
In particular, $C_{ijk} = C^{(0)}_{ijk}$ is the tree-level three-point
function, related to
the curvature of $\cM$ by the special geometry formula\footnote{Note
that our definition of $C_{ijk}$ differs from the one used in most
of the supergravity literature by a factor $e^K$.}
\be
\label{rcc}
R_{i\bar j k \bar l}= g_{i\bar j} g_{k \bar l} +
g_{i\bar l} g_{k \bar j} - e^{2K} C_{ikm} \bar C_{\bar j\bar l \bar n}
g^{m\bar n}.
\ee
The covariant derivatives in \eqref{chiral-C} incorporate the Levi-Civita
connection acting on $\cT^*$, given in coordinates by
$\che \Gamma_{ij}^k = g^{k\bar k} \pa_i g_{j\bar k}$,
as well as the connection $\che -\pa K$ in $\cL$.
So, for example,
\begin{align} \label{dfgc}
\che D_i  F_g &= (\pa_i - (2g-2) \pa_i K) F_g, \\
\che D_i  C_{j}^{(g)} &= ( \pa_i  - (2g-2) \pa_i K -
\Gamma_{ij}^k \pa_k ) C_{j}^{(g)}.
\end{align}
In \cite{Bershadsky:1994cx} the topological string partition function is
written
\be
\label{psibcov}
\che \Psi_{\rm BCOV} = \exp\left[ W(t^i,\bar t^{\bar j}; x^i, \lambda) \right].
\ee
Because of the last term in \eqref{BCOVW},
$\Psi_{\rm BCOV}$ is a section of $\cL^{1-\frac{\chi}{24}}$.\footnote{More
precisely, $\Psi_{\rm BCOV}$ is a section of the line bundle obtained by pulling back $\cL^{1-\frac{\chi}{24}}$
from $\cM$; we will abuse language in this way several times.}
The holomorphic anomaly equations can be expressed as
two conditions on $\Psi_{\rm BCOV}$:
\begin{gather}
\che \left[\dwrt{\bar t^i} -
\frac{\lambda^2}{2} e^{2K} \bar C_{\bar i\bar j\bar k} g^{j\bar j} g^{k\bar k}
\frac{\pa^2}{\pa x^j\pa x^k}
+ g_{\bar i j} x^j \left(
\lambda \frac{\pa}{\pa \lambda} + x^k \frac{\pa}{\pa x^k} \right) \right] \Psi_{\rm BCOV} = 0,
\label{BCOV1}\\
\che \left[\dwrt{t^i} - \Gamma_{ij}^k x^j \frac{\pa}{\pa x^k}
+ \pa_i K \left( \frac{\chi}{24}-1 - \lambda \frac{\pa}{\pa\lambda} \right)
- \frac{\pa}{\pa x^i} + \pa_i F_1 + \frac{1}{2\lambda^2}
C_{ijk} x^j x^k\right] \Psi_{\rm BCOV} = 0.\label{BCOV2}
\end{gather}
The second equation \eqref{BCOV2} summarizes the relations \eqref{BCOVW}, \eqref{chiral-C},
while \eqref{BCOV1} reproduces the holomorphic anomaly equations proper, which relate
the antiholomorphic derivative of the genus-$g$ correlation function
to correlators at lower genera.

\subsection{Verlinde's holomorphic anomaly equations} \label{sec-verlinde}

E.~Verlinde's version of the holomorphic anomaly
equations \cite{Verlinde:2004ck} is
obtained from \eqref{BCOV1}, \eqref{BCOV2} by making two transformations.
The first step
is to rescale $x^i_{\rm new} = \lambda^{-1} x^i_{\rm old}$,
so that the first term in \eqref{BCOVW} vanishes at $\lambda=0$.
After this rescaling $x^i$ are coordinates in $\che \cT \otimes \cL$, so
altogether $(t^i,\bar t^\bi,x^i,\lambda^{-1})$
are coordinates on the total space of
$\cL \oplus (\cT \otimes \cL)$ over $\cM$; we call this total space $\ctM$.
Second, note that the anomaly equation for the one-loop
vacuum amplitude \cite{Bershadsky:1993ta},
\be
\label{1loopan}
\che \dwrt{t^i} \dwrt{\bar t^{\bar j}} F_1 = \half e^{2K}
C_{ikl} \bar C_{\bar j\bar k \bar l} \ g^{k\bar k} g^{l\bar l}
-\left( \frac{\chi}{24} -1 \right) g_{i\bar j},
\ee
(which is contained in \eqref{BCOV1} at leading order in $\lambda$),
is solved locally by
\footnote{This equation corrects a sign error in
(4.7) of \cite{Verlinde:2004ck}, which affects (4.4) and (4.5).}
\be \label{local-sol}
F_1 = - \half \log |g| + \left( \frac{n_v+1}{2}
- \frac{\chi}{24} + 1 \right) K + f_1(t) + \bar f_1(\bar t).
\ee
Here $f_1(t)$ is a holomorphic function
and $|g|=\det(g_{i\bar j})$.
To patch these local solutions together
into the global $F_1$, $e^{f_1(t)}$ should be the coordinate expression of a section of
$\che \cL^{\frac{\chi}{24}-1 - \frac{n_v + 1}{2}} \otimes \cK_\cM^{\half}$, where $\cK_\cM$ is the canonical bundle of $\cM$, locally trivialized by the section $dt^1 \wedge \cdots \wedge dt^{n_v}$.
Then defining
\be
\label{psiv}
\che \psiv(t,\bar t;x,\lambda) = e^{f_1(t)}
\Psi_{\rm BCOV}(t,\bar t; \lambda x, \lambda),
\ee
$\psiv$ is a section of $\cL^{-\frac{n_v+1}{2}} \otimes \cK_\cM^{\half}$ over $\ctM$, and
the anomaly equations
for $\psiv$ are expressed purely in terms of the special geometry of
$\cM$:
\begin{gather}
\che \left[ \dwrt{\bar t^i} -
\half e^{2K} \bar C_{\bar i\bar j\bar k} g^{j\bar j} g^{k\bar k}
\frac{\pa^2}{\pa x^j\pa x^k}
- g_{\bar i j} x^j \frac{\pa}{\pa \lambda^{-1}} \right] \psiv = 0, \label{Ver1b}\\
\che \left[ \nabla_{t^i} - \Gamma_{ij}^k x^j \frac{\pa}{\pa x^k} -
\half \partial_{t^i} \log |g| -
\lambda^{-1}\frac{\pa}{\pa x^i} + \half C_{ijk} x^j x^k \right] \psiv = 0,
\label{Ver2b}
\end{gather}
where
\be
\che \nabla_i = \dwrt{t^i} + \pa_i K
\left( x^k \frac{\pa}{\pa x^k} + \lambda^{-1} \frac{\pa}{\pa{\lambda^{-1}}}
+ \frac{n_v+1}{2} \right).
\ee
A crucial observation for what follows \cite{Verlinde:2004ck} is that,
provided both $\psiv$ and $\Psi'_{\mathrm V}$ obey \eqref{Ver1b} and \eqref{Ver2b},
\be
\label{vinner}
\begin{split}
\che \int dx^i ~d\bar x^{\bar i}~ & d\lambda^{-1} ~d\bar \lambda^{-1}~
\sqrt{|g|} ~  e^{-\frac{n_v+1}{2}K} \\
&\exp\left( -e^{-K} x^i g_{i\bar j} \bar x^j + e^{-K}
\lambda^{-1}\bar\lambda^{-1} \right)
\psiv^{'*}(t,\bar t;\bar x,\bar\lambda) ~ \psiv(t,\bar t;x,\lambda)
\end{split}
\ee
is a pure number, independent of $(t,\bt)$.  This is proven by formal integration
by parts.

\subsection{The wave function property and coherent states}
\label{sec-wave-function}

The holomorphic anomaly equations were elegantly reinterpreted in \cite{Witten:1993ed} as the
wave function property of the topological string.  In this section
we review this reinterpretation.

Recall that $\psiv$ is defined on the total space $\ctM$ of
the holomorphic vector bundle $\cL \oplus (\cT \otimes \cL)$ over $\cM$.
The fiber at $t \in \cM$ of this vector bundle can be identified with the fixed
vector space $H^3(X,\R)$, equipped with a particular ``Griffiths'' complex structure $\cJ_t$.
Explicitly, the correspondence between an element $\gamma \in H^3(X,\R)$
and the coordinates $(\lambda^{-1}, x)$
on $\cL \oplus (\cT \otimes \cL)$ is given by the Hodge decomposition of $H^3(X,\C)$:
\begin{equation}
\label{hodge}
\begin{array}{ccccccccc}
\gamma &=& \lambda^{-1} \Omega &+& x^i \delta_i \Omega &+&  x^{\bar i} \delta_{\bar i}
\bar \Omega &+& \bar\lambda^{-1} \bar\Omega \\
&\in& H^{3,0} &\oplus& H^{2,1} &\oplus& H^{1,2} &\oplus& H^{0,3} \\
&\simeq& \cL &\oplus& (\cT \otimes \cL) &\oplus& (\bcT \otimes \bcL) &\oplus& \bcL.
\end{array}
\end{equation}
Here we introduced $\delta_i \Omega = (\pa_i + \pa_i K) \Omega$, which is the $(2,1)$ part of $\pa_i \Omega$.
Using this decomposition one can define
\begin{equation}
\cJ_{t} = \left\{ \begin{array}{ccc}
+\I & \mbox{on} & H^3_+ = H^{3,0} \oplus H^{2,1}, \\
-\I & \mbox{on} & H^3_- = H^{0,3} \oplus H^{1,2}.
\end{array} \right.
\end{equation}
Then $H^3(X,\R)$ with the complex structure $\cJ_{t}$
is identified with the $+\I$-eigenspace $H^3_+$.  So as a complex vector bundle over $\cM$,
$\cL \oplus (\cT \otimes \cL)$ is isomorphic to the bundle of $+\I$-eigenspaces $H^3_+$,
which we write as $\cH^3_+$.

$H^3(X, \R)$ also carries a canonical
symplectic structure, inherited from the antisymmetric pairing
$(\alpha,\beta)=\int_X \alpha \wedge \beta$ on $\Omega^3(X,\IR)$:  in our
$(x,\lambda^{-1})$ coordinates it is
\be
\label{sympl}
\che \omega = \I~ e^{-K} ~ \left( g_{i\bar j} dx^i \wedge
d\bar x^{\bar j} - d\lambda^{-1} \wedge d\bar\lambda^{-1} \right).
\ee
This $\omega$ is compatible with $\cJ_{t}$ for every $t$, so
the ${\cal J}_{t}$ give holomorphic polarizations on
the fixed space $H^3(X,\R)$.  Witten's proposal \cite{Witten:1993ed},
revisited in \cite{Verlinde:2004ck}, is that
$\psiv(t,\bar{t}; x, \lambda)$ at fixed $t$ should be viewed as a
wave function obtained by quantizing $H^3(X,\R)$ in this holomorphic
polarization,\footnote{It should be stressed that,
despite the adjective ``holomorphic'', $\psiv$ does depend on $\bar t$;
the holomorphy here refers to the dependence on the phase space
coordinates $(x,\lambda)$ only; the coordinates $(t,\bar t)$ are not part of the
phase space we are quantizing, but rather parameterize the possible polarizations.}
and the anomaly equations \eqref{Ver1b},
\eqref{Ver2b} simply implement the infinitesimal Bogoliubov
transformation which results from a change of
$(t,\bar{t})$.

In other words:  there is some Hilbert space
$\cH$ obtained by quantizing $H^3(X,\R)$ with the symplectic form $\omega$.
In $\cH$ there should be a state $\ket{\Psi}$ which contains the
background-independent information in the topological string theory.
Further, there should be a $(t,\bar t)$-dependent
basis of $\cH^*$ consisting of ``coherent states'' $_{(t, \bar{t})} \!\langle x,\lambda |$,
such that
\be
\label{psicoh}
\che \psiv(t, \bar{t}; x, \lambda) = \ _{(t, \bar{t})} \!\langle
 x,\lambda | \Psi\rangle.
\ee
The holomorphic anomaly equations obeyed by $\psiv$ then express the variation
with $(t, \bt)$ of the states $_{(t, \bar{t})} \!\langle x,\lambda |$; they give
no constraint on the state $|\Psi\rangle$.

Let us prove a few facts about the basis of coherent states.
By definition they diagonalize the $(t,\bar t)$-dependent
annihilation operators $\wh{\lambda^{-1}}$ and $\wh{x^i}$:
\begin{equation}
_{(t, \bar{t})} \!\langle x,\lambda |\  \wh {\lambda^{-1}} (t,\bar t) =
_{(t, \bar{t})} \!\langle x,\lambda |\           {\lambda^{-1}}, \quad
_{(t, \bar{t})} \!\langle x,\lambda |\  \wh{x^{i}}(t, \bt) =
_{(t, \bar{t})} \!\langle x,\lambda |\  x^i.
\end{equation}
We also want to work out the action of the
creation operators $\wh{\bar \lambda^{-1}} (t,\bar t)$
and $\wh{\bar x^{\bar i}}(t,\bar t)$.  These should be the adjoints of the annihilation operators.
To determine them explicitly we need to use the formula \eqref{vinner}, now interpreted as
\label{vinner-as-ip}
\begin{multline}
\che \langle \Psi' | \Psi\rangle =
\int dx^i ~d\bar x^{\bar i}~ d\lambda^{-1} ~d\bar \lambda^{-1}~
\sqrt{|g|} ~  e^{-\frac{n_v+1}{2}K} \\
\exp\left( -e^{-K} x^i g_{i\bar j} \bar x^{\bar j} + e^{-K}
\lambda^{-1}\bar\lambda^{-1} \right)
\IP{\Psi'|\bar x, \bar \lambda} \IP{x,\lambda|\Psi},
\end{multline}
which expresses the ``completeness relation'' for the basis of coherent states,
\be
\label{vinnerc}
\che \int dx^i ~d\bar x^{\bar i}~  d\lambda^{-1} ~d\bar \lambda^{-1}~
\sqrt{|g|} ~  e^{-\frac{n_v+1}{2}K}
~\exp\left( -e^{-K} x^i g_{i\bar j} \bar x^{\bar j} + e^{-K}
\lambda^{-1}\bar\lambda^{-1} \right)
|\bar x,\bar\lambda\rangle \langle x,\lambda| = {\mathbf 1}.
\ee
Using \eqref{vinnerc} and integrating by parts, we can show that
\begin{equation} \label{adjoints}
\che \cbra \wh{\bar \lambda^{-1}} = -e^{K} \dwrt{\lambda^{-1}} \cbra, \quad \cbra \wh{\bar x^{\bi}} = e^{K} g^{\bi j} \dwrt{x^j} \cbra.
\end{equation}
consistently with the commutation rules which follow from \eqref{sympl},
\be
[\wh{\lambda^{-1}},\wh{\bar \lambda^{-1}}]=e^{K}\ ,\quad
[\wh{x^{i}},\wh{\bar x^{\bar j}}] = - e^{K} g^{i\bar j}
\ee
Then exponentiating these derivative operators, one can construct all of the coherent
states by acting with creation operators on a fixed vacuum:
\begin{equation} \label{coherent-build-b}
\che \cbra = _{(t,\bt)}\!\langle \Omega \rvert \exp\left[ e^{-K}
\left( - \lambda^{-1} \wh{\bar\lambda^{-1}}
+ x^i g_{i\bar j} \wh{\bar x^{\bar j}} \right) \right],
\end{equation}
where $\bra{\Omega} = \bra{x^i = 0, \lambda^{-1} = 0}$.

Substituting \eqref{coherent-build-b} in \eqref{psicoh}, we conclude that
the topological wave function in Verlinde's presentation can be
written as an overlap
\be
\label{overlap-b}
\che \psiv(t, \bar{t}; x, \lambda) = \ _{(t, \bar{t})} \!\langle
 \Omega |
\exp\left[ e^{-K}
\left( - \lambda^{-1} \wh{\bar\lambda^{-1}}
+ x^i g_{i\bar j} \wh{\bar x^{\bar j}} \right) \right]~
| \Psi\rangle.
\ee
To reduce clutter, we will sometimes
drop the explicit $(t,\bar t)$ dependence of coherent states and creation/annihilation operators from the notation.

We conclude this section with two comments.
First, note that the Hermitian metric constructed from $\omega$ and $\cJ_t$ is not
positive definite, but rather has signature $(n_v,1)$
(as reflected by the wrong-sign Gaussian in the integrand of
\eqref{vinner}).  This point, well known in the mathematics literature
and also noted \ti{e.g.} in \cite{Dijkgraaf:2002ac},
implies that the coherent states we considered above are
not normalizable states.
Hence it seems that these coherent states must be considered as somewhat formal objects.
To perform the quantization rigorously in this indefinite-signature polarization,
one should perhaps introduce higher cohomology instead of trying to represent the states
as holomorphic functions.  Alternatively, one could switch to a polarization for which the metric
is positive definite.  There is a natural choice available, namely
the ``Weil'' complex structure $\cJ'_{t}$, given by the Hodge $\star$ operator
on $\Omega^3(X,\R)$, which acts as
\begin{equation}
\label{weilj}
\che \cJ'_{t} = \left\{ \begin{array}{ccc}
+\I & \mbox{on} & H^{3,0} \oplus H^{1,2}, \\
-\I & \mbox{on} & H^{0,3} \oplus H^{2,1}.
\end{array} \right.
\end{equation}
Formally, one can transform from the $\cJ_t$ polarization to the $\cJ'_t$ polarization
by Fourier transforming over $\lambda^{-1}$ (see Appendix B).
One complication in this approach is that $\cJ'_{t}$ does not vary
holomorphically
with $t$ (unlike $\cJ_t$.)  Throughout this section we work formally in the indefinite-signature polarization,
the one in which we can make direct contact with the anomaly equations.

Second, let us comment on the occurrence of the line bundle
$\cL^{- \frac{n_v+1}{2}} \otimes\cK_\cM^{\half}$ pulled back from $\cM$,
where both $\psiv$
and the coherent states $_{(t, \bar{t})} \!\langle x,\lambda |$
are valued.  This bundle, which we call $\cK^\half$, is naturally isomorphic to a square root of
$\bigwedge^{n_v + 1} \left( \cT^* \otimes \cL^* \oplus \cL^* \right)$.  In other words,
an element of $\cK^\half$ is the same as a square-root of a holomorphic top-form on
$H_+^3$.  The wave
function $\psiv$ is hence naturally ``twisted'' by the bundle of half-forms.
Such a twisting makes no essential difference if one
looks only at a fixed $t$, but it introduces phase factors in the Bogoliubov
transformations which arise upon varying $t$; indeed, twisting by $\cK^\half$
is exactly what is needed to ensure that the wave
function has trivial monodromy upon traveling around a contractible loop in $\cM$.
This is known as the ``metaplectic correction'' to the quantization of
the phase space (see \eg \cite{MR94a:58082,gt-parallel}), and plays an important r\^ole in making the topological
amplitude covariant under electric-magnetic duality.

\subsection{Homogeneous coordinates} \label{sec-projective}

To simplify the anomaly equations further, it is convenient to replace $\cM$ by
the space $\chM$ of complex structures on $X$ with a chosen
holomorphic 3-form $\Omega$.
The benefit of so doing is that $\chM$ admits a class of particularly nice
coordinate systems.  To construct them, first note that $\chM$ can be
embedded into $H^3(X,\C)$ just by mapping a point of $\chM$ to its corresponding
$\Omega$.  This embedding realizes $\chM$ as a complex Lagrangian cone inside $H^3(X,\C)$.

Suppose we choose a
symplectic basis (or ``marking'') $\{A^I, B_I\}$, $I = 0, \dots, n_v$, for $H_3(X,\R)$.  Then define
coordinates on $H^3(X,\C)$ by
\begin{equation}
\che X^I = \int_{A^I} \Omega,\quad F_I = \int_{B_I} \Omega,
\end{equation}
and introduce the ``characteristic function''
\begin{equation}
\quad F = \half X^I F_I.
\end{equation}
The cone $\chM$ is said to be ``generated'' in this basis by
$F(X)$:  this means $\chM$ is the subset of $H^3(X,\C)$ defined by
\begin{equation}
\che F_I(X) = \frac{\pa F}{\pa X^I}.
\end{equation}
In particular, a point $(X^I, F_I)$ on $\chM$
is determined locally
by its $X^I$, so the $X^I$ give a complex coordinate system on $\chM$, which can also be viewed
as a homogeneous coordinate system on $\cM$.
We also introduce
\begin{align}
\che \tau_{IJ}(X) &= \pa_I \pa_J F, \\
\che C_{IJK}(X) &= \pa_I \pa_J \pa_K F. \label{def-c}
\end{align}
In what follows it will be convenient to think of $\psiv$ as defined on $\chM$;
concretely, we consider $\psiv$ to depend on the $X^I$ as
independent variables, and to be invariant under their overall rescaling.

We expand $\gamma \in H^3(X,\R)$ in terms of the
basis $\{\partial_I \Omega\}$ instead of
the basis $\{\Omega, \delta_i \Omega\}$, writing
\begin{equation} \label{hodge-large}
\che \gamma = \half \left( e^{\qtr \pi \I} x^I \partial_I \Omega + e^{- \qtr \pi \I} \bar{x}^I \bar{\partial}_I \bar{\Omega} \right),
\end{equation}
where the phase was inserted for later convenience.
These coordinates $x^I$ on $H^3(X,\R)$ are related to the $(x^i, \lambda^{-1})$ of \eqref{hodge} by
\be
\label{lphase}
\che \half e^{\qtr \pi \I} x^I = \lambda^{-1} X^I + x^i D_i X^I,
\ee
where we introduced
\be
\label{deffh}
\che D_i X^I = \int_{A^I} \delta_i \Omega = (\pa_i + \pa_i K) X^I.
\ee
The $x^I$ are sometimes called ``large phase space coordinates''.

The symplectic form on $H^3(X,\R)$ becomes in these coordinates
\be
\label{sympl2}
\che \omega = \ihalf~[\im\tau]_{IJ} ~ dx^I \wedge d\bar x^J.
\ee
This leads, upon quantization, to the commutators
\be
\label{comxxb}
\left[\wh{x^I}, \wh{\bar x^J}\right] = -2 [\im\tau]^{IJ}.
\ee
Since the change of variable \eqref{lphase} is holomorphic (at fixed $X$),
there is no Bogoliubov transformation required in rewriting
the wave function $\psiv$ in the $x^I$ variables.
In other words, the coherent states $\cbra$ which diagonalize $\wh{\lambda}$ and $\wh{x^i}$
also diagonalize $\wh{x^I}$.  However, we now choose a different normalization:
namely we write
\be
_{(X,\bar{X})}\!\langle \Omega \rvert = \sqrt{J} \, _{(t,\bar{t})}\!\langle \Omega \rvert
\ee
where $J$ is the Jacobian of the change of coordinates \eqref{lphase}.

Substituting the new coordinates in \eqref{adjoints} gives the action of
operators on coherent states,
\be
\label{cohlps}
\che \bcohb \wh{x^I} = \bcohb x^I,\quad
\bcohb \wh{\bx^J} = 2 [\im\tau]^{JK} \frac{\pa}{\pa x^K} \bcohb.
\ee
Similarly, substituting in \eqref{coherent-build-b} gives\footnote{We are suppressing the $\bar{X}$
dependence of $_X\!\langle \Omega \rvert$ to lighten the notation.}
\begin{equation} \label{coherent-build-v}
\che \bcohb = \,_X\!\langle \Omega |
\exp\left[ \half x^I [\im\tau]_{IJ}  \wh{\bar x^J} \right].
\end{equation}
Thus the overlap formula for $\psiv$, \eqref{overlap-b}, may be rewritten as
\be
\label{topoverx}
\che \psiv(X^I, \bar{X}^I; x^I) =\,_X\!\langle \Omega | \exp\left[ \half x^I [\im\tau]_{IJ} \wh{\bar x^J} \right] | \Psi\rangle.
\ee
Here we have rescaled $\psiv$ by a factor of $\sqrt{J}$.  Geometrically we can describe this rescaling as follows:
having introduced the large phase space coordinates we have a natural section $\sqrt{\prod_I dx^I}$ of $\cK^\half$,
and we are now writing $\psiv$ with respect to this trivialization, which differs from the one we used previously
by the factor $\sqrt{J}$.\footnote{Since this explicit
rescaling factor may be unexpected we make two further comments about it.  First, $J$ could have been
absorbed in a redefinition of $f_1$, which would amount to replacing \eqref{local-sol} by
\begin{equation*}
F_1 = - \half \log \det[\im\tau] + \left(- \frac{\chi}{24} + 1 \right) K + f_1(t) + \bar f_1(\bar t).
\end{equation*}
Second, if we choose the coordinates $t^i = X^i / X^0$ and a corresponding
canonical \kahler potential $K$, then $J$ is just a numerical constant.
}

Now we are ready to summarize the properties of $\psiv$ in this new coordinate system.
First, using the easy fact
\begin{equation} \label{jac-formula}
\abs{J}^2 = \frac{\det[\im\tau]}{|g|~e^{-(n_v+1)K}},
\end{equation}
the inner product \eqref{vinner} becomes
\be \label{inner-2}
\che \langle \Psi' | \Psi\rangle =
\int dx^I~ d\bar x^I ~\sqrt{\det[\im\tau]}~
e^{-\half x^I [\im\tau]_{IJ} \bar x^J} ~ \psiv^{'*}(\bar x^I) \psiv(x^I).
\ee
Finally, the anomaly equations \eqref{Ver1b}, \eqref{Ver2b}
become \cite{Verlinde:2004ck,Dijkgraaf:2002ac}
\begin{gather}
\che \left[\frac{\pa}{\pa \bar X^I} + \ihalf\bar C_I^{JK}
\frac{\pa^2}{\pa x^J\pa x^K}\right] \psiv = 0,\label{Ver1c}\\
\che \left[ \frac{\pa}{\pa X^I}
- \half \frac{\pa}{\pa X^I} \log\det[\im\tau]
+ \ihalf C_{IJ}^K x^J \frac{\pa}{\pa x^K}
+ \ieighths C_{IJK} x^J x^K \right] \psiv = 0, \label{Ver2c}
\end{gather}
where the indices are raised and lowered with the metric
$[\im\tau]_{IJ}$ and its inverse $[\im\tau]^{IJ}$.

\subsection{The real polarized topological amplitude}

An obvious nuisance in representing $\ket{\Psi}$ in the holomorphic polarizations given by $\cJ_t$ is
the dependence on the
basepoint $t\in\cM$, which as we have reviewed is responsible for the holomorphic anomaly equations.
An alternative (emphasized in \cite{Verlinde:2004ck}) is to consider instead a real polarization, based on the
decomposition of $\gamma\in H^3(X,\R)$ with respect to a symplectic basis $(A^I,B_I)$ of $H_3(X,\R)$,
with dual basis $(\alpha_I, \beta^I)$:
\be
\label{realdec}
\che \gamma = p^I \alpha_I + q_I \beta^I.
\ee
This gives coordinates $(p^I, q_I)$ on $H^3(X,\R)$ which do
not require a choice of $t$; we have traded that dependence for the
dependence on the chosen symplectic basis.
The relation between the decomposition \eqref{realdec} and the one considered above in \eqref{hodge-large} is
\be
\che p^I = \re(x^I),\quad q_I = \re(\tau_{IJ} x^J);
\ee
so upon quantization, $p^I$ and $q_I$ should become Hermitian operators, with
\be
\wh{p^I} = \re( \wh{x^I}),\quad
\wh{q_I} = \re( \tau_{IJ} \wh{ x^J}),
\ee
while the symplectic form $\omega=dq_I\wedge dp^I$ gives
\begin{equation}
[\wh{p^I},\wh{q_J}]=\I \delta^I_J.
\end{equation}
So now we can introduce coherent states $\ket{p^I}$ with
\be
\wh{p^I} |p^I\rangle= p^I |p^I\rangle,\quad
\wh{q_I} |p^J\rangle=  \I \frac{\pa}{\pa p^I}
|p^J\rangle,
\ee
normalized to obey the completeness relation
\begin{equation}
\int\ dp^I\ \ket{p^I} \bra{p^I} = {\mathbf 1}.
\end{equation}
If we write $\ket{\Omega}_\R = \ket{p^I = 0}$, then the
coherent states can be constructed from $\ket{\Omega}_\R$ by
\begin{equation} \label{coherent-build-r}
\ket{p^I} = \exp (-\I p^I \wh{q_I}) \ket{\Omega}_\R.
\end{equation}
Using these coherent states one can represent $\ket{\Psi}$ in a real polarization:
\begin{equation}
\Psi_{\IR}(p^I) = \langle p^I | \Psi \rangle.
\end{equation}
This $\Psi_\R$ is related to
$\psiv(X,\bar X; x) =\,_{(X^I,\bar X^I)}\!\langle x^I |\Psi\rangle$ by a Bogoliubov transformation,
\be
\label{psixpsip}
\psiv(X,\bar X; x) = \int dp^I~ _{(X^I,\bar X^I)}\!\langle x^I | p^I \rangle \Psi_{\IR}(p^I).
\ee
To write this transformation explicitly we need to determine the intertwining function
$_{(X^I,\bar X^I)}\!\langle x^I | p^I \rangle$.  This is the holomorphically polarized
form of the state $\ket{p^I}$, so we call it $\psiv(p^I; X,\bar X; x)$.
Substituting the expressions \eqref{coherent-build-v}, \eqref{coherent-build-r} for the coherent states, we have
\begin{equation}
\psiv(p^I; X,\bar X; x)\ =\ _X\!\IP{ \Omega | \exp \left(\half x^J [\im \tau]_{JK} \wh{\bx^K}\right) \exp\left(-\I p^I \wh{q_I}\right) | \Omega }_\R.
\end{equation}
Then using $\widehat{\bx^I} = 2 \wh{p^I} - \wh{x^I}$, $\widehat{q_I} = \bar\tau_{IJ} \widehat{p^J} + \I [\im \tau]_{IJ} \widehat{x^I}$ and commuting the
operators $\wh{x^I}$, $\wh{p^J}$ (such
that $[\wh{x^I}, \wh{p^J}]=-\im\tau^{IJ}$)
to annihilate the
vacua at the two ends gives
\be
\label{intert-1}
\psiv(p^I; X,\bar X; x)\ =\ \left(_X\!\IP{ \Omega | \Omega }_\R \right) \exp\left[ -\ihalf p^I \bar\tau_{IJ} p^J
+ p^I [\im\tau]_{IJ} x^J
- \qtr x^I [\im\tau]_{IJ} x^J  \right].
\ee
The normalization factor $_X\!\IP{ \Omega | \Omega }_\R$ can be determined (up to an unimportant
overall constant) by requiring that
$\psiv(p^I; X,\bar X; x)$ obeys the equations \eqref{Ver1c}, \eqref{Ver2c}:  it turns out to be
$\sqrt{\det [\im \tau]}$, so altogether\footnote{This
function was first obtained in \cite{Dijkgraaf:2002ac},
by factorizing the semiclassical
partition function of the NS5-brane wrapped on $X$.}
\begin{equation}
\label{intert}
\psiv(p^I; X,\bar X; x)\ =\ \sqrt{\det [\im \tau]} \exp\left[
-\ihalf p^I \bar\tau_{IJ} p^J
+ p^I [\im\tau]_{IJ} x^J
- \qtr x^I [\im\tau]_{IJ} x^J  \right].
\end{equation}
In Appendix A, we discuss how the real polarization
relates to the ``holomorphic
ambiguities'' which appear in Section 7 of \cite{Bershadsky:1993ta}.

We close with one comment.  Above we have emphasized that the real polarization
involves the choice of a symplectic basis of
3-cycles (symplectic marking).  As
usual in quantum mechanics, the wave functions $\Psi_\R$, $\Psi_{\R'}$ in two different symplectic bases
differ by a Bogoliubov transformation:  \eg under $B_I\to B_I+m_{IJ} A^J$, $\Psi_\R$
transforms by
\be
\Psi_{\IR'}(p^I) = e^{\I p^I m_{IJ} p^J} ~\Psi_{\IR}(p^I),
\ee
and under an exchange $A_I \leftrightarrow B^I$,
\be
\label{four}
\Psi_{\IR'}(q_I) = \int dp^I e^{\I p^I q_I} \Psi_\R(p^I).
\ee
There is a small subtlety here:  it turns out that in order to fix some
sign ambiguities in the Bogoliubov transformations one needs to consider the wave
function as a half-density.  So it is $\Psi_{\IR}(p^I) \sqrt{\abs{\prod dp^I}}$
that is actually well defined \cite{MR94a:58082}.  This is
another aspect of the metaplectic correction which we
mentioned in Section \ref{sec-wave-function}.

\subsection{The (anti)holomorphic topological
amplitude} \label{sec-holomorphic}

So far we have mainly reviewed forms of the topological string partition function already
found in the literature.  In the next two subsections we introduce two new variants.

We begin with the holomorphically polarized topological
amplitude \eqref{topoverx},
\be
\label{topoverx-replay}
\psiv(X^I, \bar{X}^I; x^I) = \ _X\!\langle
\Omega | \exp\left[ \half x^I [\im\tau]_{IJ} \wh{\bar x^J} \right]
| \Psi\rangle.
\ee
As we have discussed, this $\psiv$ is a section of the holomorphic line bundle $\cK^\half$ over the
complex manifold $\ctM$, the total space of $\cH^3_+$ over $\cM$.  It obeys
holomorphic anomaly equations which are close to, but not identical to, those satisfied by the complex
conjugate of a theta function.  The obvious difference is that
a theta function $\theta(t,z)$
is purely holomorphic both in $t$ and in $z$.  So we want to make some transformation so that
$\psiv$ will become purely antiholomorphic.

To do so, we need to overcome three obstacles, two of which are apparent even without looking at the
detailed form of the anomaly equations.
First, $\cK^\half$ is a holomorphic line bundle over $\ctM$,
so there is no well defined notion of an antiholomorphic section.  Second, $\psiv$ depends holomorphically rather
than antiholomorphically on the fiber coordinates $x^I$.
Both of these problems can be solved using the Hermitian metric in $\cH^3_+$
(which in our coordinates is $[\im \tau]_{IJ}$.)  First,
using this metric one can identify $\cK^\half \simeq \bar{\cK}^{-\half}$, so that we will have a section of
an antiholomorphic line bundle instead of a holomorphic one.  In our coordinates and local trivialization
this just means dividing by $\sqrt{\det [\im \tau]}$.
Second, one can also identify
the total space of $\cH^3_+$ with that of $(\bar \cH^3_+)^* = (\cH^3_-)^*$, thus
reversing the complex structure on the fibers, so that $\psiv$ will be antiholomorphic on them.  This
amounts to the change of coordinate $x^I = [\im \tau]^{IJ} \bar y_J$.

Making these transformations in \eqref{topoverx-replay} leads to the function
\begin{equation} \label{firsttry}
\ _X \!\langle
\Omega_\ahol |\exp\left[ \half \bar y_J \wh{\bar x^J} \right]| \Psi\rangle.
\end{equation}
Here we have rescaled $\langle \Omega\rvert = \sqrt{\det[\im\tau]} \langle \Omega_\ahol\rvert$.
Still, \eqref{firsttry} cannot be antiholomorphic in $X$:  the trouble now is with the
annihilation operator $\wh{\bar x^J}$, which we recall depends implicitly on $X$.
But since $\wh{x^J}$ annihilates $\langle \Omega_\ahol|$ one may replace
$\wh{\bar x^J}$ by $\wh{x^J}+\wh{\bar x^J}=2 \wh{p^J}$, at the expense
of introducing an extra phase from the Baker-Campbell-Hausdorff formula in commuting
$\wh{x^J}$ past $\wh{\bar x^J}$:
\begin{equation} \label{secondtry}
_X\!\langle \Omega_\ahol |\exp\left[ \bar y_J \wh{p^J} \right]| \Psi\rangle = \exp\left( \qtr x^I [\im\tau]_{IJ} x^J \right) \ _X \!\langle \Omega_\ahol |\exp\left[ \half \bar y_J \wh{\bar x^J} \right]| \Psi\rangle.
\end{equation}
The left side of \eqref{secondtry} is our candidate for the
antiholomorphic version of the topological amplitude:
\be
\label{topovery}
\Psi_\ahol(X^I, \bar{X}^I; \bar y_I) =
\ _X \!\langle
\Omega_\ahol | \exp\left[\bar y_I \wh{p^I} \right]
| \Psi\rangle.
\ee
In terms of the original $\psiv$, we can summarize our transformation as
\begin{equation}
\label{psih}
\Psi_\ahol(X^I,\bar X^I; \bar y_I) =
\frac{\exp\left( \qtr x^I [\im\tau]_{IJ} x^J \right)}{\sqrt{\det[\im\tau]}}
\psiv(X^I, \bar X^I; x^J).
\end{equation}
Now we want to see what sort of anomaly equations $\Psi_\ahol$ obeys.
The simplest way to work this out is to apply the transformation
\eqref{psih} to the explicit solutions \eqref{intert} giving the intertwiner to the real polarization:
they become
\be
\label{interthl}
\Psi_\ahol (p^I; X,\bar X; \bar y_I) =
\exp\left[ - \ihalf p^I \bar\tau_{IJ} p^J + p^I \bar y_I \right].
\ee
To avoid the nuisance of working with antiholomorphic objects, from now on
we use the holomorphic $\Psi_\hol = \Psi^*_\ahol$, a section of $\cK^{-\half}$.
Then in terms of $\Psi_\hol$, the anomaly equations take the simplified form\footnote{Note
that \eqref{holom-a-2}, \eqref{holom-a-1} could be summarized as
$\delta \Psi_\hol = \ihalf \delta \tau_{JK} \pa_{y_J}\pa_{y_K} \Psi_\hol$,
which is formally identical to the heat equation giving the variation of
a holomorphic Siegel theta function $\Theta(\tau,y)$ with $\tau$.  Indeed $\Psi_\hol$ behaves
very much like a Siegel theta function, except that $\tau$ is not allowed to
vary arbitrarily, and worse yet $\im \tau$ is not positive definite, so
the usual construction of holomorphic Siegel theta functions would lead to a divergent series.}
\begin{gather}
\left[\frac{\pa}{\pa X^I} - \ihalf C_{IJK}
\frac{\pa^2}{\pa y_J\pa y_K}\right] \Psi_\hol = 0, \label{holom-a-2} \\
\frac{\pa}{\pa \bar{X}^I} \Psi_\hol = 0. \label{holom-a-1}
\end{gather}
The inner product \eqref{inner-2} of wave functions becomes
\be \label{inner-3}
\langle \Psi' | \Psi\rangle =
\int dy_I~d\bar y_I~
\sqrt{\det[\im\tau]}~
e^{-\half (y_I+\bar y_I)[\im\tau^{-1}]^{IJ}(y_J+\bar y_J)}
(\Psi'_\hol)^*(\bar X^I; \bar y_I) \Psi_\hol(X^I; y_I).
\ee

\subsection{The generalized heat equation}

For comparison to the results we will find in Section \ref{sec-symmetric} we will
make one further transformation of the partition function.  This transformation can
be made only after choosing a privileged element $A^0$ in $H_3(X,\R)$, with corresponding period
$\int_{A^0} \Omega = X^0$.  Having done so, one may define coordinates on $\cM$ by
\begin{equation}
t^i = \frac{X^i}{X^0}.
\end{equation}
Now we make a small further adjustment:  we divide the coordinates $y_I$ into $\{y_0, y_i\}$,
and then replace $y_0$ by
\begin{equation} \label{affine-shift}
w = y_0 + t^i y_i.
\end{equation}
(This shift will find a natural interpretation in Section \ref{sec-sym-anom}.)
Considering $\Psi_\hol$ as a function of $(t^i, y_i, w)$,
\eqref{holom-a-2}, \eqref{holom-a-1}
then become
\begin{gather}
\left[ \dwrt{t^{i}} - \ihalf C_{ijk} \frac{\partial^2}{\pa y_{j} \pa y_{k}} + y_{i} \dwrt{w} \right] \Psi_\hol = 0, \label{heat} \\
\dwrt{\bar t^i} \Psi_\hol = 0. \label{purehol}
\end{gather}
This is not too hard to check using the relation $X^I C_{IJK} = 0$, which follows from the definition \eqref{def-c} and the
homogeneity of $F$, and which in these coordinates implies
$p^J p^K C_{iJK} = (p^j - p^0 t^j)(p^k - p^0 t^k) C_{ijk}$.

Altogether then, we have found that the complex conjugate of the BCOV topological string
amplitude can be transformed into a strictly holomorphic function $\Psi_\hol$, obeying a heat equation
analogous to that of the Jacobi theta series.  To be precise, we gave two closely related versions of this heat equation,
namely \eqref{holom-a-2} and \eqref{heat}.

The existence of such a heat equation is not unexpected:  indeed, it was the similarity between the
anomaly equation of \cite{Bershadsky:1994cx} and the heat equation that led Witten to propose the wave function
interpretation of the topological string partition function.  This interpretation was natural particularly in light of the earlier work
\cite{Axelrod:1991xt}, in which the heat equation for (abelian and non-abelian) theta functions was identified as a
``background dependence'' in the quantization of the moduli space of flat connections on a Riemann surface,
coming from the need to choose a complex structure.  However, the
analogy has always been slightly puzzling:  if the topological string partition function arose in this way one would
expect it to be purely holomorphic, as was the case for the theta functions of \cite{Axelrod:1991xt}.  We view
the transformations relating $\Psi^*_\BCOV$ to $\Psi_\hol$, given above, as the resolution of this small puzzle:
in summary, $\Psi^*_\BCOV$ can be made holomorphic by choosing the proper complex structure
on its domain, and by choosing a different convention for the genus zero $0$,$1$,$2$-point functions than
was chosen in \cite{Bershadsky:1994cx}.

In Section \ref{sec-symmetric}, we will give a representation-theoretic construction of the anomaly
equations for $\Psi_\hol$, exhibiting its wave function nature directly, in cases where $\cM$ is
a symmetric space.

\subsection{Modularity} \label{sec-modular-general}

So far we have considered the topological string partition function on the \tm space
$\cM$ and its extension to $\ctM$.  In the string theory, though, the relevant space is generally not
$\cM$ but some quotient $\Gamma \backslash \cM$, where $\Gamma$ is a ``modular group'' of large biholomorphisms
of $X$.  So we should ask how the topological string partition function
transforms under $\Gamma$.

For the original partition function $\Psi_{\rm BCOV}$ the answer is simple.  Since $\Psi_{\rm BCOV}$ is defined
directly in terms of the B model topological string theory, which appears to be $\Gamma$-invariant, $\Psi_{\rm BCOV}$
should also be $\Gamma$-invariant.  However, we must be careful about what this means, since $\Psi_{\rm BCOV}$
is a section of $\cL^{1 - \frac{\chi}{24}}$.  So to be precise, the action of $\Gamma$ on $X$ induces
holomorphic actions on $\cM$, $\cT$ and $\cL$, and we expect that $\Psi_{\rm BCOV}$ is invariant under the combined
action.  For the modified partition function $\psiv$ the story is similar, except that $\psiv$ is a section of
$\cK^\half$ instead of $\cL^{1 - \frac{\chi}{24}}$.

If we choose some specific coordinates on $H^3(X,\C)$ and trivialization of $\cK^\half$, then we can be more concrete.
For example, suppose we choose the coordinates $x^I$ and trivialization of $\cK^\half$ by the half-form
$\sqrt{\prod_I dx^I}$, as in Section \ref{sec-projective}.  An element $\gamma \in \Gamma$
acts on $H^3(X,\C)$ by some element in $Sp(2n_v+2,\Z)$, which we also write as $\gamma$:  in a basis where
the symplectic form is the block-diagonal $\bigl( \begin{smallmatrix} 0 & 1 \\ -1 & 0 \end{smallmatrix}
\bigr)$, write
\begin{equation}
\gamma = \begin{pmatrix} A^I_J & B_{IJ} \\ C^{IJ} & D^I_J \end{pmatrix}.
\end{equation}
Based on the above we expect that $\psiv(t,x) \sqrt{\prod_I dx^I}$ should be invariant under the action of $\gamma$.
This leads to the transformation law
\begin{equation} \label{modular-psiv}
\left(\det [C \tau + D] \right)^{\half} \psiv\left(\gamma t, (C \tau + D) x\right) = \psiv(t, x).
\end{equation}
The point $\gamma t$ can be characterized by the fact that
the period matrix $\tau_{IJ}$ transforms as $\gamma\tau=
(A\tau+B)(C\tau+D)^{-1}$.

Now what about $\Psi_\hol$?
The transformation \eqref{psih} relating $\psiv$ to $\Psi^*_\hol$ introduces extra factors
which depend explicitly on the choice of a symplectic basis.
With these factors included, the transformation law \eqref{modular-psiv} is modified to
\begin{equation} \label{modular-hol}
 \left(\det [C \tau + D]\right)^{-\half} \Psi_\hol\left(\gamma t, (C \tau + D)^{-1} y\right) = \exp \left(-\ihalf y_I \left[(C\tau + D\right)^{-1}]^I_J
C^{JK} y_K\right) \Psi_\hol(t, y),
\end{equation}
just as for the Riemann theta series \cite{MR688651}.

This modular property in principle gives strong constraints on $\Psi_\hol$.
One concrete way to extract these constraints would be to make a perturbative expansion of
$\log \Psi_\hol$.  At each genus we would then expect to find a holomorphic section of a
holomorphic line bundle over $\Gamma \backslash \cM$.  If $\Gamma \backslash \cM$ can be
compactified while keeping control of the behavior near the boundary points,
then such a line bundle has a finite-dimensional space of holomorphic sections; in other words, $\Psi_\hol$
is determined up to a finite-dimensional ambiguity at each genus.  This seems to be consistent
with the results of \cite{Huang:2006si,Aganagic:2006wq}, who have recently studied
the constraints modular invariance and the anomaly equation impose on topological string perturbation theory.

\section{The holomorphic anomaly for symmetric tube domains} \label{sec-symmetric}

Now we consider a special class of $\N=2$, $d=4$ supergravity theories coupled to
vector multiplets.  These theories are constructed from Euclidean Jordan algebras $J$
with invariant cubic norms, and can be
obtained by dimensional reduction from corresponding theories in
five dimensions, which were constructed in
\cite{Gunaydin:1983rk,Gunaydin:1983bi,Gunaydin:1984ak}. In the
five-dimensional description, the vector fields are in 1-1
correspondence with basis elements of $J$. After dimensional
reduction this correspondence is extended: both electric and
magnetic field strengths get related to basis elements of the
``Freudenthal triple system'' defined over $J$
\cite{Ferrara:1997uz,Gunaydin:2000xr,Gunaydin:2005gd,Gunaydin:2005zz,Bellucci:2006xz}.

Given $J$ we review the construction of this Freudenthal triple
system as well as a special K\"ahler manifold $\cM$, which is the
vector multiplet moduli space of the corresponding $d=4$
supergravity theory, and a global symmetry group $G$ which acts
transitively on $\cM$ by isometries.  $\cM$ is equipped with a
privileged set of special coordinates, in which the prepotential is
directly related to the cubic norm of $J$.

Most of these
supergravity theories are not known to arise as low energy limits of
Type II string theory compactified on a Calabi-Yau
threefold.  One notable exception comes from Type II string theory
compactified on a quotient of $K3 \times T^2$, considered in
\cite{Ferrara:1995yx} and recently reinvestigated in
\cite{Klemm:2005pd,Maulik:2006cf}.  The vector multiplet moduli
space of this theory has a classical $SO(10,2)\times SU(1,1)$
symmetry acting transitively. \ti{A priori} this symmetry could be
broken by genus zero worldsheet instantons, but it turns out that
these instantons are absent, so the exact special geometry of the
moduli space is $SO(10,2) \times SU(1,1)$ symmetric.  Indeed, this
model gives a string theory realization of one of the generic Jordan
family of $\N=2$ Maxwell-Einstein supergravity theories, with $n=11$
vector multiplets (the first row of Figure \ref{table} below),
coupled to 12 hypermultiplets.

It would be interesting to have more examples of magical supergravity
theories which can be embedded into string theory.  In fact, several
candidate theories can be found in \cite{Sen:1995ff}, among them
a model which at low energy describes the
coupling of 15 vector multiplets to $\N=2$ supergravity.  It is easy
to show that, at least before considering worldsheet instantons, this is the ``magical''
supergravity theory with moduli space $SO^*(12) / (SU(6) \times U(1))$ (the fifth row
in Figure \ref{table}).  It is not known whether the $SO^*(12)$
symmetry of this moduli space survives into the quantum theory.

In any case, the holomorphic anomaly equation can be constructed from an $\N=2$, $d=4$ supergravity theory
without reference to its embedding into string theory.  From now on we adopt this point of view.

\subsection{The phase space}

We begin with a Euclidean Jordan algebra $J$, of real dimension $n_v$.\footnote{A review of the relevant aspects of
Jordan algebras can be found in \cite{Gunaydin:1983bi,MR0238916}; in what follows we will not need to know much about their algebraic structure.}  We assume that $J$ admits a cubic norm $N$, which may be written concretely in linear coordinates
\begin{equation} \label{def-norm}
N(p) = \sixth C_{ijk} p^i p^j p^k.
\end{equation}
From $J$ one can construct an associated ``Freudenthal
triple system'' (see \eg \cite{MR0063358,kansko,MR0225838,MR0170974,MR0248185,MR0374223,MR0295205});
this is a real vector space $V$ which decomposes as
\begin{equation} \label{freudenthal-decomp}
V = \R \oplus J \oplus J \oplus \R.
\end{equation}
We choose linear coordinates $p^0, p^i, q_i, q_0$ respectively on the four summands (using the Jordan algebra trace to
pair the two copies of $J$).  We also introduce the index $I = (0,i)$.  Then $V$ is equipped with a natural symplectic form $\omega$,
\begin{equation}
\omega = dp^0 \wedge dq_0 + dp^i \wedge dq_i = dp^I \wedge dq_I,
\end{equation}
and a quartic form constructed from the norm on $J$,
\begin{equation}
I_4 = 4 p^0 N(q_i) - 4 q_0 N(p^i) + 4 q^i_\sharp\ p_i^\sharp - (p^0 q_0 + p^i q_i)^2.
\end{equation}
where $p_i^\sharp:=\half C_{ijk}p^j p^k, q^i_\sharp:=\half C^{ijk}q_j q_k$.
We define $G$ as the group of linear transformations of $V$ which
preserve $\omega$ and $I_4$; this $G$ shows up as a global symmetry of the supergravity theory, and in particular it
acts by isometries on the moduli space $\cM$ as we will see below.

There is a particularly important generator $D \in \fg$ which recovers the
real decomposition \eqref{freudenthal-decomp},
acting with eigenvalues $-3$, $-1$, $+1$, $+3$ respectively on the four summands.
It also induces a decomposition of $\fg$ into real subspaces
\begin{equation} \label{g-grading}
\fg = \fg_- \oplus \fg_0 \oplus \fg_+
\end{equation}
where $\fg_-$, $\fg_0$, $\fg_+$ have eigenvalues $-2$, $0$, $2$ respectively.
The three subspaces can be conveniently described in terms of $J$.  Namely,
$\fg_0$ is the ``structure algebra'' of $J$,
consisting of infinitesimal linear transformations of $J$ preserving $N$ up to overall rescaling; these
act on $V$ in the obvious way.  $\fg_-$ and $\fg_+$ are isomorphic as vector spaces to $J$; we
write the generators as $S_i$ and $T^i$ respectively, and then define $R^j_i = [S_i, T^j]$.
The trace of $R$ gives the abovementioned $D$, as $D = \frac{3}{n_v} R^i_i$;
we also define the traceless part $D^i_j = R^i_j - \frac{1}{3} D$.

\subsection{The \tm space} \label{sec-moduli}

Next we need to describe the \tm space $\cM$.
As we mentioned above,
this space has been studied extensively in the physics
literature \cite{Gunaydin:1983rk,Gunaydin:1983bi,Cremmer:1984hc,Cremmer:1984hj,deWit:1991nm,deWit:1992cr,deWit:1992wf};
for a more mathematical review and pointers to the mathematical literature see \eg the book \cite{MR1446489}.
Here we present a construction of $\cM$
described in \cite{MR1758493}, which is tailored to our needs: it
mimics the procedure of Section \ref{sec-projective},
where the \tm space of a Calabi-Yau threefold $X$ was realized as
the projectivization of a Lagrangian cone $\chM$ inside the
complexified phase space $H^3(X,\C)$.

In our current setup $V$ is playing the role of $H^3(X,\R)$.
So we should begin by complexifying the phase space $V$ to $V_\C$.  Then the
action of $G$ on $V$ extends holomorphically to an action of a
complexified group $G_\C$ on $V_\C$.  Inside $V_\C$, we consider the
orbit $\cO$ of the ``lowest weight'' vector $(p^0, p^i, q_i, q_0) = (1,0,0,0)$.  From
the fact that any element of $\fg_\C$ only changes the $D$-grading by $0$ or $\pm 2$,
it follows that $\cO$ is a Lagrangian cone; the projectivization of $\cO$ is
a compact manifold which we call $\cM_D$.  This is almost, but not
quite, the desired $\cM$:  indeed, by construction $\cM_D$ is a single
orbit of $G_\C$, but under the action of the real group $G \subset G_\C$ it has a
further decomposition into orbits.  The desired $\cM$ is a ``big'' orbit (open in
$\cM_D$) for which the stabilizer of any point is a maximal compact
subgroup of $G$.  By construction,
then, $\cM \simeq G / K$ with $K$ a maximal compact subgroup of $G$.
Similarly we can write $\cM_D$ as $G_\C / P$ with $P$ the
stabilizer of $(1,0,0,0)$.  It will be useful below to remember that
$\cM$ is embedded into $\cM_D$.\footnote{To make this embedding $\cM
\subset \cM_D$ seem more familiar, we recall a closely related
example.  Consider $G_\C = SL(2,\C)$ acting on $V_\C = \C^2$.  Then
the orbit $\cO$ of the vector $(0,1)$ is $\C^2$ minus the origin.
Projectivizing gives the Riemann sphere $\C\PP^1$ as $\cM_D$.  Then
letting $SL(2,\C)$ act on the Riemann sphere by linear fractional
transformations, the stabilizer $P$ of the south pole consists of
lower-triangular matrices; so $\cM_D$ is $SL(2,\C) / P$.  On the
other hand, if one considers just the action of $SL(2,\R)$, then
$\cM_D$ breaks up into orbits, namely the two hemispheres and the
equator.  On either the top or bottom hemisphere, the stabilizer in
$SL(2,\R)$ of a point is $U(1)$; we call the top hemisphere $\cM$.  So
the standard realization of the upper half plane $\cM$ as $SL(2,\R) /
U(1)$ is naturally embedded inside the realization of the Riemann
sphere $\cM_D$ as $SL(2,\C) / P$.}
In particular, this embedding
immediately shows that $\cM$ has a complex structure on which $G$ acts
by holomorphic automorphisms.  Moreover, it gives a natural notion of
analytic continuation to the boundary of $\cM$.

This construction of $\cM$ inside $\PP(V_\C)$ gives a nice class of
homogeneous coordinate systems on $\cM$, coming from
symplectic markings of $V$, just as we discussed in Section \ref{sec-projective} for
symplectic markings of $H^3(X,\R)$.  Actually, in this case the
situation is even better: the decomposition \eqref{freudenthal-decomp}
of $V$ gives a natural symplectic marking of $V$, which turns out to
give an especially convenient projective coordinate system $X^I$.  The
prepotential $F$ (defined as the function which generates the Lagrangian cone
in these coordinates, as in Section \ref{sec-projective}) then turns out to be purely cubic,
\begin{equation}
\label{prepot}
F = \frac{N(X^i)}{X^0}.
\end{equation}
The decomposition \eqref{freudenthal-decomp} also gives a natural
1-dimensional subspace (hence a natural choice of the $0$-th direction), which one can use to define honest coordinates
on $\cM$,
\begin{equation}
t^i = \frac{X^i}{X^0}.
\end{equation}
For example, taking the Jordan algebra $J = \R$, the above
construction gives $\cM$ as the upper half-plane with its standard
coordinate $t = x + \I y$. More generally, the coordinates $t^i = x^i
+ \I y^i$ realize $\cM$ as a ``generalized (K\"ocher) upper half-plane'' inside $J_\C$, of the form \cite{Gunaydin:1983bi}
\begin{equation}
\cM = \{ t = x + \I y:  x \in J, y \in J_+ \} \subset J_\C,
\end{equation}
where $J_+$ is the cone of ``positive definite'' elements in $J$, with boundary contained in the ``cubic light-cone'' $\{y: N(y) = 0\}$.
$\cM$ also has a natural K\"ahler metric inherited from $V_\C$, with the
simple K\"ahler potential
\be
\label{kahl}
K(t^i,\bar t^\bi) = - \log N(t^i - \bar t^\bi).
\ee
In these coordinates, the action of $G$ on $\cM$ can be described simply, \eg because
it comes from the linear action on $V_\C$:  one finds that the generators
$T_i$, $R^i_j$, $S^i$
act respectively by translations,
Lorentz rotations/dilatations and ``special conformal'' transformations \cite{Gunaydin:1992zh},
\begin{equation}
T_i \mapsto \dwrt{t^i}, \quad
R^i_j \mapsto -\Lambda^{ik}_{jl} t^l \dwrt{t^k}, \quad
S^i \mapsto \half \Lambda^{ij}_{kl} t^k t^l \dwrt{t^j},
\end{equation}
where $\Lambda^{ij}_{kl} = \delta^i_k \delta^j_l + \delta^j_k
\delta^i_l - C_{klm} C^{ijm}$.\footnote{Our
$C$ differs from that of GST
\cite{Gunaydin:1983bi} by a factor $2/\sqrt{3}$,
$C^{\mathrm {here}}_{ijk}=
\frac{2}{\sqrt{3}} C_{ijk}^{\mathrm {GST}}$,
to agree with the conventions
used in the special geometry literature.}
Here we introduced $C^{ijk}$, where the indices are raised
using the metric on $\cM$; this actually gives a constant
tensor, numerically equal to $C_{ijk}$ up to an overall constant, which can be shown using the
fact that both $C^{ijk}$ and $C_{ijk}$ descend from invariant
tensors in the five-dimensional supergravity theory \cite{Gunaydin:1984ak}.

Note that the action of $G$ preserves the condition that all $t^i$ are real.  So $G$ can be thought of as
a ``generalized conformal group'' ${\mathrm{Conf}}(J)$ which acts on the Jordan algebra $J$, as well as
on $\cM \subset J_\C$ \cite{Gunaydin:1992zh}.  (This generalizes the fact that $SL(2,\R)$ acts on $\R$ as
well as on the upper half-plane.)  The action of $G$ on $J$ preserves separations along the cubic light-cone,
i.e. $N(x - x') = 0 \implies N(gx - gx') = 0$.

Now we have finished
constructing the phase space $V$, the \tm space $\cM$, and
the group $G$, which we have seen
acts naturally both on $V$ (by linear maps) and on $\cM$ (by holomorphic automorphisms).
In Figure \ref{table} we tabulate the Euclidean Jordan algebras of
degree 3 and their corresponding $G$, $K$, $V$ and $n_v$.

\begin{figure}
\begin{center}
\begin{tabular}{|c|c|c|c|c|} \hline
$J$ & $G$ & $K$ & $V$ & $n_v$ \\ \hline \hline
$\R \oplus \Gamma_{n-1,1}$ & $SU(1,1) \times SO(2,n)$ & $U(1) \times U(1) \times SO(n)$ & $\mathbf{(2,n+2)}$ & $n+1$ \\ \hline
\R & $SL(2,\R)$ & $U(1)$ & $\mathbf{4}$ & 1 \\ \hline
$Herm_3({\mathbb R})$ & $Sp(6,\R)$ & $U(1) \times SU(3)$ & $\mathbf{14}$ & 6 \\ \hline
$Herm_3({\mathbb C})$ & $SU(3,3)$ & $U(1) \times SU(3) \times SU(3)$ & $\mathbf{20}$ & 9 \\ \hline
$Herm_3({\mathbb H})$ & $SO^*(12)$ & $U(1) \times SU(6)$ & $\mathbf{32}$ & 15 \\ \hline
$Herm_3({\mathbb O})$ & $E_{7(-25)}$ & $U(1) \times E_6$ & $\mathbf{56}$ & 27 \\ \hline
\end{tabular}
\end{center}
\caption{Euclidean Jordan algebras of degree 3 and their
corresponding $G$, $K$, $V$ and $n_v$.  The groups $G$, $K$ are
given only up to a finite cover or finite quotient. $Herm_3$ denotes
$3 \times 3$ Hermitian matrices; $\Gamma_{n-1,1}$ is a Jordan
algebra of degree 2 defined by a quadratic form of Lorentzian
signature \cite{Gunaydin:1983bi}.} \label{table}
\end{figure}

\subsection{Quantization of phase space:  the Heisenberg group}

Following the discussion of Section \ref{sec-revisited}, we now want to
understand the topological string partition function in these models as a state $\ket{\Psi}$ in the quantization of $V$.
This quantization as usual gives a Hilbert space $\cH_\hbar$ acted on by operators corresponding to the coordinates
$\{p^I, q_I\} \in V^*$.  These operators together with a single central generator $Z$ generate a Heisenberg
group $H$, with the standard commutation relations\footnote{We write the commutation relations for $\I \wh{p}$ and $\I \wh{q}$
rather than $\wh{p}$ and $\wh{q}$ because we want to deal with the skew-Hermitian generators of the real group $H$.}
\begin{equation}
[\I\wh{p^I}, \I\wh{q_J}] = - \delta^I_J Z,
\end{equation}
where $Z$ acts on $\cH_\hbar$ by a pure imaginary scalar $\I \hbar$.  In what follows we will always set $\hbar = 1$; this choice is consistent with Section \ref{sec-revisited} where we used the commutation relations
$[\wh{p^I},\wh{q_J}] = \I \delta^I_J$.

To construct a ``topological string partition function'' obeying the holomorphic anomaly equations
from the abstract state $\ket{\Psi}$, we want to introduce a basis of coherent states for each point
$t \in \cM$.  Hence we need to define a complex structure $\cJ_t$ on $V$.  We define it as in Section \ref{sec-wave-function}, in
terms of a Hodge decomposition.
Recall that $D$ gave a real decomposition of $V$, shown in \eqref{freudenthal-decomp}.  More generally, its conjugate
$D_t = (\Ad(\exp t^i T_i)) D$ gives a conjugate decomposition
\begin{equation}
V_\C \, = \, V_{3} \oplus V_{1} \oplus V_{-1} \oplus V_{-3}
\end{equation}
where for complex $t$ the eigenspaces are not invariant under conjugation.  This is our desired Hodge
decomposition at the point $t \in \cM_D$, from which $\cJ_t$ can be constructed as in Section \ref{sec-wave-function}.  This Hodge decomposition has been discussed previously in \cite{MR1258484}.

\subsection{The Schr\"odinger-Weil representation}

So far we have discussed the action of the Heisenberg group $H$ on the Hilbert space $\cH_\hbar$.  In the present case
there is another natural group around, namely $G$, which acts on $V$ by symplectic transformations.
The symplectic transformations induce an action on $\cH_\hbar$:  $g \in G$ gives a Bogoliubov transformation $B(g)$,
satisfying the condition\footnote{The existence of some $B(g)$ satisfying \eqref{bog-defined} can be abstractly
proven by the following standard argument:  for a fixed $g$, we consider two different actions of $H$ on $\cH_\hbar$, one by
$\I f \mapsto \I\wh{f}$, the other by $\I f \mapsto \I\wh{gf}$.  Because $g$ preserves the symplectic structure, these are both representations of $H$, with the same central character $Z \mapsto \I \hbar$.  Hence the
Stone-von Neumann Theorem guarantees that they are equivalent, so the desired $B(g)$ exists.}
\begin{equation} \label{bog-defined}
B(g) \wh{f} B(g)^{-1} = \wh{gf}.
\end{equation}
The requirement \eqref{bog-defined}, together with unitarity, is enough to determine $B(g)$ up to a phase ambiguity.  In
particular, it implies that $B(g) B(g') = e^{\I \theta(g, g')} B(gg')$.  One can then try to choose the ambiguous phases
in such a way that $B(g) B(g') = B(gg')$.  This is not possible in general, but one can always manage it by passing
to a ``metaplectic'' covering group of $G$; in what follows we will implicitly replace $G$ by this cover when necessary.

So we have actions of $G$ and $H$ on the Hilbert space $\cH_\hbar$, with the commutation between the two determined by \eqref{bog-defined}; hence we can think of $\cH_\hbar$ as a representation of an extended group $\tG = G \ltimes H$.
Following \cite{MR1634977} we call it the ``Schr\"odinger-Weil representation''.
At least formally, it has a natural extension to a representation of a complexified group
$\tG_\C = G_\C \ltimes H_\C$, acting holomorphically on $\cH_\hbar$; in the construction of the
topological amplitude below, we work formally with the complexified generators.

Let us describe $\tG$ in a bit more detail.
The Lie algebra $\tfg$
has a decomposition which naturally extends \eqref{g-grading}:  in addition to the grading by the dilatation generator
$D$ which runs from $-3$ to $3$, there is a second grading $\Delta$ (not implemented by any element of $\tfg$, but see
Section \ref{sec-3d})
under which $\fg$ has grade 0, $V^*$ grade 1, and $Z$ grade 2.  We represent the
two gradings in Figure \ref{fj-picture}.
\begin{figure}
\begin{center}
\begin{tabular*}{0.5\textwidth}{@{\extracolsep{\fill}}c|ccccccc}
& $-3$ & $-2$ & $-1$ & $0$ & $1$ & $2$ & $3$ \\ \hline
2 & & & & $Z$ & & & \\
1 & $\I\wh{q_0}$ & & $\I\wh{q_i}$ & & $\I\wh{p^i}$ & & $\I\wh{p^0}$ \\
0 & & $S^i$ & & $R^i_j, D$ & & $T_i$ &
\end{tabular*}
\end{center}
\caption{A list of the generators of the Lie algebra $\tfg$, with two gradings, by $D$ (horizontal)
and by $\Delta$ (vertical).}\label{fj-picture}
\end{figure}
Moreover, we can realize the Schr\"odinger-Weil representation $\cH_\hbar$ (with $\hbar = 1$)
concretely in the real polarization given by our chosen symplectic marking:  then $\cH_\hbar$ just consists of $L^2$ functions of the $n_v+1$ real variables $p^I$, and the generators of $\tfg$ act by
\begin{gather}
Z \mapsto \I, \label{swrep-1} \\
\I\hq_0 \mapsto \dwrt{p^0}, \quad \I\wh{q_i} \mapsto \dwrt{p^i}, \quad \I\wh{p^i} \mapsto \I p^i, \quad \I\wh{p^0} \mapsto \I p^0,   \label{swrep-2} \\ \displaybreak[0]
S^i \mapsto - \ihalf C^{ijk} \frac{\pa^2}{\pa p^j \pa p^k} + p^i \dwrt{p^0}, \quad T_i \mapsto + \ihalf C_{ijk} p^j p^k - p^0 \dwrt{p^i}, \label{swrep-3} \\
R^j_i \mapsto \delta^j_i p^0 \dwrt{p^0} - p^j \dwrt{p^i} + \half C_{ikl} C^{jnl} \left(p^k \dwrt{p^n} + \dwrt{p^n} p^k\right). \label{swrep-4}
\end{gather}
Now we recall that $C$, constructed from the norm of the Jordan algebra $J$ in \eqref{def-norm},
satisfies a crucial ``adjoint identity'' \cite{Gunaydin:1983bi} which implies
\begin{equation}
C_{ikl} C^{inl} = \left(\frac{n_v}{3} + 1\right) \delta^n_k.
\end{equation}
Then defining $D =
\frac{3}{n_v} R^i_i$ and using \eqref{swrep-4} we find
\begin{equation}
D \mapsto 3 p^0 \dwrt{p^0} + p^i \dwrt{p^i} + \half (n_v+3). \label{swrep-5}
\end{equation}
For later use, we note a few facts which can be easily checked using this explicit realization.

First, we introduce a state $\ket{\Omega}_0$ defined by the constant function $f(p) = 1$.\footnote{This $\ket{\Omega}_0$ is not normalizable,
so it does not strictly speaking belong to $\cH_\hbar$.}  This state is annihilated
by the generators $\wh{q^I}$; these generators are not the annihilation operators of any holomorphic polarization, but it turns out that
they arise as the $t=0$ limit of the annihilation operators of $\cJ_t$, so we can sensibly refer to $\ket{\Omega}_0$ as a
``vacuum at $t=0$.''  Now we note that $\ket{\Omega}_0$ is also annihilated by $S^i$ and $D^i_j$, and has eigenvalue $\half (n_v+3)$ under $D$.

Second, we note that the operator equation
\begin{equation} \label{op-anom-0}
Z T_i + \half C_{ijk} \wh{p^j} \wh{p^k} - \wh{p^0} \wh{q_i} = 0
\end{equation}
holds in the \sw representation.  In the next section we will identify \eqref{op-anom-0} with the holomorphic anomaly equation.

\subsection{The anomaly equations} \label{sec-sym-anom}

In this section we will argue that $\Psi_\hol$, when considered as a function of both $t$ and $y$, has a natural
interpretation as a matrix element of the \sw representation, and that if one adopts this point of view from the outset,
the anomaly equations arise naturally from the operator equations \eqref{op-anom-0}.

First recall from \eqref{topovery} that for fixed $t \in \cM$,
$\Psi_\hol(t, y)$ can be understood as a matrix element of $H$ acting on the Hilbert space $\cH_\hbar$,
between $\ket{\Psi}$ and a vacuum $\ket{\Omega}_t$ in the $\cJ_t$ polarization:
\begin{equation} \label{me-heis}
\Psi_\hol(t, y) = \IP{\Psi | \exp (y_I \wh{p^I}) | \Omega}_t.
\end{equation}
On the other hand, as we have just seen, the Bogoliubov transformations which relate the vacua
in various polarizations can be implemented by the group $G_\C$ acting in the \sw representation.  In particular,
vacua in the $\cJ_t$ polarization are related to those in the $\cJ_{t'}$ polarization by the action of any
$g \in G_\C$ which maps $t$ to $t'$ in $\cM$.
Now recall that $\ket{\Omega}_0$ is a vacuum in the polarization at $t=0$ and
that $T_i$ act by translations on $\cM_D$, so $\exp \left(t^i T_i \right)$ maps $t=0$ to $t$.
With this in mind, we write
\begin{equation} \label{def-vac}
\ket{\Omega}_t = \exp \left(t^i T_i \right) \ket{\Omega}_0.
\end{equation}
Since $\ket{\Omega}_0$ is annihilated by the special conformal generators $S_i$ and transforms irreducibly
under the ``Lorentz'' generators $R_i$ with definite scale dimension under $D$, one can think of
$\Psi_\hol$ as a generalized conformal wave function associated with a unitary representation of $G$
in the sense of \cite{Gunaydin:1999jb}.

One could ask whether \eqref{def-vac} actually defines the same vacuum that appears in \eqref{me-heis}.
After all, the vacuum is only determined up to rescaling, so \ti{a priori} the choice \eqref{def-vac} could
differ by a function of $(t,\bt)$ from the one that should appear in \eqref{me-heis}.  We will verify that
\eqref{def-vac} is the correct choice by showing that
\begin{equation} \label{psi-gt}
\Psi_\hol(t^i, y_I) = \IP{\Psi | \exp (y_I \wh{p^I}) \exp \left(t^i T_i \right) | \Omega}_0
\end{equation}
obeys exactly the holomorphic anomaly equations \eqref{holom-a-2}, \eqref{holom-a-1} (which would
certainly be spoiled if one multiplied $\Psi_\hol$ by some non-constant function of $(t,\bt)$.)
Indeed, taking \eqref{psi-gt} as the definition of $\Psi_\hol$, the holomorphic anomaly equations
follow directly from facts about the Schr\"odinger-Weil representation of $\tG$, as we now show.

It is convenient to begin by reversing the order of the operators that appear in \eqref{psi-gt}.
Using $[\hp^j, T_i] = \delta^j_i \hp^0$ and $[\hp^0, T_i] = 0$, \eqref{psi-gt} becomes
\begin{equation} \label{psi-reversed}
\Psi_\hol(t^i, y_I) = \IP{\Psi | \exp \left(t^i T_i \right) \exp \left(y_i \wh{p^i} + (y_0 + t^i y_i) \wh{p^0} \right) | \Omega}_0.
\end{equation}
So setting $w = y_0 + t^i y_i$,
\begin{equation} \label{psi-rev-2}
\Psi_\hol(t^i, y_i, w) = \IP{\Psi | \exp \left(t^i T_i \right) \exp \left(y_i \wh{p^i} + w \wh{p^0} \right) | \Omega}_0.
\end{equation}
Note that this shift from $y_0$ to $w$ is exactly the one made in \eqref{affine-shift}, which converted
the anomaly equations \eqref{holom-a-2}, \eqref{holom-a-1} to their shifted variants
\eqref{heat}, \eqref{purehol}.  We will now verify that \eqref{psi-rev-2} satisfies \eqref{heat}, \eqref{purehol},
which we reproduce for convenience:
\begin{gather}
\left[ \dwrt{t^{i}} - \ihalf C_{ijk} \frac{\partial^2}{\pa y_{j} \pa y_{k}} + y_{i} \dwrt{w} \right] \Psi_\hol = 0, \label{heat-repeat} \\
\dwrt{\bar t^i} \Psi_\hol = 0. \label{purehol-repeat}
\end{gather}
The second anomaly equation \eqref{purehol-repeat} just says
$\Psi_\hol$ is holomorphic in $t^i$; this is manifest from \eqref{psi-gt}, since $\tG_\C$ acts holomorphically.
To establish the first anomaly equation \eqref{heat-repeat}
we use the operator equation \eqref{op-anom-0} which holds in the Schr\"odinger-Weil representation.
Inserting this operator into the right side of \eqref{psi-reversed} gives
\begin{equation}  \label{anom-op-1}
\IP{\Psi |  \exp \left(t^j T_j \right)
\left( Z T_i + \half C_{ijk} \wh{p^j} \wh{p^k} - \wh{p^0} \wh{q_i} \right)  \exp \left(y_j \wh{p^j} + w \wh{p^0} \right) | \Omega}_0 = 0.
\end{equation}
The $\wh{q_i}$ operator annihilates $\ket{\Omega}_0$; commuting it through $\exp y_i \wh{p^i}$ gives a commutator $[\wh{q_i}, y_i \wh{p^i}] = -y_i Z$.
Also, recall that $Z$ acts as the scalar $\I \hbar = \I$.  Altogether \eqref{anom-op-1} becomes
\begin{equation}  \label{anom-op-2}
\IP{\Psi |  \exp \left(t^j T_j \right)  \left(\I T_i + \half C_{ijk} \wh{p^j} \wh{p^k} + \I \wh{p^0} y_i \right)  \exp \left(y_j \wh{p^j} + w \wh{p^0} \right) | \Omega}_0 = 0,
\end{equation}
which is simply
\begin{equation}
\left[ \I \dwrt{t^i} + \half C_{ijk} \frac{\pa^2}{\pa y_j \pa y_k} + \I y_i \dwrt{w} \right] \Psi_\hol = 0.
\end{equation}
This is our desired anomaly equation \eqref{heat-repeat}.

\subsection{Geometry of the matrix elements}

In the last section we realized $\Psi_\hol$ as a matrix element for the group $\tG$, or more precisely its
complexification $\tG_\C$.  We now make a few extra comments about this construction, to elucidate
its geometric nature.  In particular, we will explain how the metaplectic correction we discussed in Section
\ref{sec-wave-function} arises here as a ``zero point energy'' of the vacuum $\ket{\Omega}_0$.
We will also see that the wave function is naturally
considered as a section of a prequantum line bundle, as usual in geometric quantization.

We begin by noting that our construction \eqref{psi-gt} has the general form
\begin{equation}
\Psi_\hol(t,y) = \IP{ \Psi | g_{t,y} | \Omega }_0
\end{equation}
where $g_{t,y}$ is an element in $\tG_\C$ depending holomorphically on $(t,y)$.  It is natural then to consider
$\Psi_\hol$ simply as a function on $\tG_\C$,
\begin{equation}
\Psi_\hol(g) = \IP{ \Psi | g | \Omega }_0.
\end{equation}
One might think that this function contains more information than did $\Psi_\hol(t,y)$, since the number of
variables in $g$
is more than in $(t,y)$.  However, $\Psi_\hol(g)$ obeys some extra constraints, arising from
the fact that $\ket{\Omega}_0$ is an eigenvector for some of the generators
of $\tG_\C$.  First, $\ket{\Omega}_0$ is annihilated by $\hq_I$, $S^i$, and $D^i_j$; letting $\tP$ denote the complex
group they generate, this implies immediately that
\begin{equation}
\Psi_\hol(g p) = \Psi_\hol(g), \quad p \in \tP.
\end{equation}
Hence $\Psi_\hol$ can be considered as a function on $\tG_\C / \tP$.  This agrees well with the interpretation
of $\Psi_\hol$ as a ``conformal wave function.''
Moreover, under the dilatation generator $D$, $\ket{\Omega}_0$ has eigenvalue $\half (n_v+3)$,
and under the central generator $Z$ of $H$, $\ket{\Omega}_0$ has eigenvalue $\I\hbar = \I$.  This implies that
\begin{equation} \label{bundle}
\Psi_\hol\left(g \exp (\lambda D + \zeta Z) p \right) = \exp \left(\half (n_v+3) \lambda + \I \zeta\right) \Psi_\hol(g), \quad \lambda \in \C,\,\zeta \in \C,\,p \in \tP.
\end{equation}
Thus one can also divide out by the larger complex group $\tQ$ generated by $\tP$, $D$ and $Z$, at the cost that
we now describe $\Psi_\hol(g)$ as a holomorphic section of a holomorphic line bundle over $\tG_\C / \tQ$,
rather than as an honest function.
This line bundle is constructed in a tautological way:  the state $\ket{\Omega}_0$ transforms in a 1-dimensional
representation of $\tQ$; call this representation $\rho$, and then define
\begin{equation}
\cV\,=\,\left\{ (g, z) \in \tG_\C \times \C \right\}\,\Big/\,\left[(g,z) \sim (gq, \rho(q) z): q \in \tQ\right].
\end{equation}
Any $\Psi_\hol$ satisfying \eqref{bundle} gives a section of $\cV$.
So our construction has naturally led to a section $\Psi_\hol$ of the line bundle $\cV$ over $\tG_\C / \tQ$.

What is the meaning of this quotient of complex groups?  Recall from Section \ref{sec-moduli}
that in defining the moduli space
$\cM$ we first considered a larger space $\cM_D$ which was a homogeneous space for $G_\C$, and then obtained
$\cM$ as an open $G$-orbit inside
$\cM_D$.  Here we have done something precisely analogous, now for the extended group $\tG$ instead of $G$:
letting $\ctM_D$ denote
$\tG_\C / \tQ$, we can construct $\ctM \subset \ctM_D$ as an open orbit for the real group $\tG$.  This $\ctM$ is the
exact counterpart of the $\ctM$ which appeared in Section \ref{sec-verlinde}:
it is a vector bundle over $\cM = G/K$, with phase-space fibers
generated by the action of $H$.  It is equipped with a natural complex structure inherited from $\ctM_D$, which varies
along the fibers, and also inherits the holomorphic line bundle $\cV$, of which $\Psi_\hol$ is a section.
It can also be written as a quotient $\ctM = \tG / \tK$, where $\tK$ is the group generated by $K \subset G$ and
the Heisenberg center $Z$ (so $\tK$ is non-compact).

How should we interpret the fact that $\Psi_\hol$ comes out as a section of the holomorphic line bundle $\cV$?
In fact, this is exactly what
one expects from the point of view of geometric quantization:  the wave functions are sections of the
``prequantum line bundle'' over the phase space.  The fact that the central generator $Z$ of
the Heisenberg algebra acts nontrivially means that the restriction of $\Psi_\hol(t,y)$ to any
fixed $t \in \cM$ is naturally a section of a line bundle over the phase space.
Of course, since the phase space here is just a
linear space, one can suppress the issue by trivializing the line bundle, which is what was done in
Section \ref{sec-revisited}.  Nevertheless, it is gratifying to see the line bundle coming out naturally here.

Similarly, the fact that the dilatation generator $D \in \fg$ acts nontrivially means that even after fixing $y=0$,
$\Psi_\hol(t)$ is still a section of a homogeneous line bundle over $\cM$,
characterized by the $D$-eigenvalue $\half(n_v+3)$.
This line bundle is exactly the bundle $\cK^{-\half}$ related to
the metaplectic correction to geometric quantization, which we discussed in Section \ref{sec-wave-function}.
To see this, it suffices to note that a holomorphic top-form on $V$ is an element of $\wedge^{\mathrm{top}}(V^*_+)$, and
$V^*_+$ has $n_v$ basis elements with weight $1$ and one basis element with weight $3$ under $\Ad(g(t)) D$,
so their wedge product has total weight $n_v + 3$.

To sum up:  We have seen that starting from a state $\ket{\Psi}$ in the \sw representation of $\tG$ one can
canonically construct a matrix element $\IP{\Psi | g | \Omega}$, which is naturally considered as a section of a holomorphic line
bundle $\cV$ over a coset space $\tG / \tK$.  This $\cV$ is the line bundle in which a wave function should live:  it
incorporates the prequantum line bundle and the metaplectic correction.
The formula \eqref{psi-gt} can be understood as defining a particular trivialization of $\cV$,
in which the holomorphic anomaly equations of Section \ref{sec-holomorphic} hold.

\subsection{Modularity} \label{sec-modular-symmetric}

It is natural to ask what happens if $\ket{\Psi}$ is invariant under some discrete group $\Gamma \subset G$.
The line bundle $\cV$ over $\tG / \tK$ defined in the last section has an obvious action of $G$.  So
we can define a quotient bundle over the double coset space $\Gamma \backslash \tG / \tK$, and
the matrix element $\IP{\Psi | g | \Omega}$ is then a holomorphic section of that quotient.  This is
the usual situation for holomorphic modular forms:  they can be understood as sections of holomorphic
line bundles over double coset spaces.  The double coset space $\Gamma \backslash \tG / \tK$ of
interest here is a vector bundle over $\Gamma \backslash G / K = \Gamma \backslash \cM$; each fiber is
a copy of the linear phase space.

If we work with a specific coordinate system on $\tG$ and trivialization of $\cV$,
such as the ones used in \eqref{psi-gt} to make contact with Section \ref{sec-revisited},
then $\IP{\Psi | g | \Omega}$ will be represented by a holomorphic function, and the $\Gamma$-invariance will imply
some modular transformation law for this function.  Starting from \eqref{psi-gt}, for example, one can derive the
transformation under $\gamma \in \Gamma$ by considering
\begin{equation}
\Psi_\hol(t^i, y_I) = \IP{\Psi | \gamma \exp (y_I \wh{p^I}) \exp \left(t^i T_i \right) | \Omega}_0
\end{equation}
and then commuting $\gamma$ to the right.  The result of this procedure is the same modular transformation law
we found in Section \ref{sec-modular-general} for the topological string partition function.
In other words, the condition that the string theory is invariant under $\Gamma$ acting by biholomorphisms
on the Calabi-Yau threefold is equivalent to the
condition that the state $\ket{\Psi}$ is invariant under $\Gamma$ acting in the \sw representation.

One could go further and ask what happens if $\Gamma$ also includes
a discrete subgroup of $H$, for example, the subgroup generated by $\exp 2 \pi \I \hp$ and $\exp 2 \pi \I \hq$.
In this case the phase-space fibers of $\Gamma \backslash \tG / \tK$ become tori, varying in a family over the moduli space.
Formally then, our construction would lead to a theta function, a holomorphic section of $\cV$ over this family of tori.  However,
this cannot be quite right:  $\cV$ actually has no holomorphic sections over the torus, because its curvature
is not positive definite but rather has one negative eigenvalue; this is a manifestation of the problem we mentioned at the end of Section
\ref{sec-wave-function}.\footnote{One might ask where exactly we went wrong.  Presumably
the issue is that in constructing $\Psi_\hol$ by the expression \eqref{psi-gt}
we are trying to analytically continue the matrix elements of the \sw representation beyond their actual domain of analyticity.  A symptom of this is that the states $\ket{\Omega}_t$ which we
use are badly non-normalizable.}
From the general theory of geometric quantization one can guess what kind of ``theta function'' one could hope for:  it should
be an element in the sheaf cohomology $H^1$ of $\cV$ rather than $H^0$.  Such theta functions have not been discussed very
much in the literature, but see \eg \cite{MR1987784, MR2084583}.  It would be interesting to understand the relation between our
formal construction (which seems to agree with the topological string) and this sheaf cohomology.  Of course, \ti{a priori} there is
no reason why the topological string should be invariant under any subgroup of $H$; we discuss this further in Section \ref{sec-discussion}.

\subsection{Fourier-Jacobi coefficients and the minimal representation} \label{sec-3d}

To this point, there has been a perfect analogy between the topological amplitude
on a Hermitian symmetric tube domain and a Jacobi theta series:
in either case, the function satisfies a holomorphic
heat equation (\eqref{heatth} or \eqref{heat}), and can be written as a
matrix element in the \sw representation of a non-reductive group
$\tG= G \ltimes H$, as
\begin{equation}
\Psi(g) = \IP{\Psi \vert g \vert \Omega}_0.
\end{equation}
Moreover, in each case the heat equations
obeyed by $\Psi(g)$ follow directly from operator equations
\eqref{sweq}, \eqref{op-anom-0} which hold in the
\sw representation.

It is natural to ask whether this analogy can be pursued further.  As discussed
in the introduction, the
\sw representation of $SL(2,\R) \ltimes H$
occurs inside the metaplectic representation
of a larger group $Sp(4,\R)$, the annihilator of which
contains the operator equation \eqref{sweq}. Can one similarly
obtain the \sw representation of $\tG$
and the operator equation \eqref{op-anom-0}
from a representation of a larger reductive group $G'$?

There is a natural candidate for this $G'$ which is
well motivated from physics:  upon reducing $\N=2, d=4$
supergravity on a circle to three dimensions, the vector multiplet moduli space $\cM$
is enlarged to a quaternionic-K\"ahler space known as the ``c-map'' of
$\cM$.  When $\cM$ is a Hermitian
symmetric tube domain $G/K$ of complex dimension $n_v$,
its c-map is a quaternionic-K\"ahler symmetric space $G' / (G^c \times SU(2))$,
of real dimension $4n_v+4$.  Here $G^c$ is a compact real form
of $G$.  We tabulate the groups in question in Table \ref{table-3d}.
\begin{figure}
\begin{center}
\begin{tabular}{|c|c|c|c|c|} \hline
$G$ & $K$ & $G'$ & $K'$ & $n_v$ \\ \hline \hline
$SU(1,1) \times$ & \multirow{2}{*}{$U(1) \times U(1) \times SO(n)$} & \multirow{2}{*}{$SO(n+2,4)$} & $SU(2) \times SU(2)$ & \multirow{2}{*}{$n+1$} \\
$SO(2,n)$ &  & & $\times SO(n+2)$ &  \\ \hline
$SL(2,\R)$ & $U(1)$ & $G_{2(2)}$ & $SU(2) \times SU(2)$ & 1 \\ \hline
$Sp(6,\R)$ & $U(1) \times SU(3)$ & $F_{4(4)}$ & $SU(2) \times Sp(6)$ & 6 \\\hline
$SU(3,3)$ & $U(1) \times SU(3) \times SU(3)$ & $E_{6(2)}$ & $SU(2) \times SU(6)$ & 9 \\ \hline
$SO^*(12)$ & $U(1) \times SU(6)$ & $E_{7(-5)}$ & $SU(2) \times SO(12)$ & 15 \\ \hline
$E_{7(-25)}$ & $U(1) \times E_6$ & $E_{8(-24)}$ & $SU(2) \times E_{7}$ & 27 \\ \hline
\end{tabular}
\end{center}
\caption{Isometry groups $G'$ of the moduli spaces occurring upon dimensional reduction from $4$ to $3$ dimensions,
with their maximal compact subgroups $K'$, and the corresponding $G$, $K$ from $4$ dimensions.  All groups are
given only up to finite covering or finite quotient.} \label{table-3d}
\end{figure}

Just as $G$ was the ``conformal group'' associated to a Jordan
algebra $J$, the group $G'$ can be described as the ``quasi-conformal
group'' associated to the same $J$ \cite{Gunaydin:2000xr,Gunaydin:2005zz}.  The reason for this
terminology is that $G'$ can be characterized as the invariance
group of the ``quartic light-cone'' in $V_\C$,
\be
\label{qlc}
I_4(p^I-\bar p^I, q_I-\bar q_I) + 2
( k - \bar k + p^I \bar q_I - \bar p^I q_I)^2 = 0,
\ee
more precisely, $G'$ is the group leaving the left side of \eqref{qlc} invariant up to
a multiplication by a function of $(p^I,q_I,k)$ times a function
of $(\bar p^I, \bar q_I,\bar k)$. The twistor space of the
quaternionic-K\"ahler space $G'/G^c\times SU(2)$
admits a Einstein-\kahler metric whose \kahler potential
can be shown to be proportional to the logarithm of \eqref{qlc},
making the symmetry under $G'$ manifest \cite{gnppw-to-appear}.

The Lie algebra $\fg'$ has a bigrading extending those of $\fg$ and $\tfg$, with a corresponding
picture extending Figure \ref{fj-picture}, given in Figure \ref{g3-picture}.
\begin{figure}
\begin{center}
\begin{tabular*}{0.6\textwidth}{@{\extracolsep{\fill}}c|ccccccc}
& $-3$ & $-2$ & $-1$ & $0$ & $1$ & $2$ & $3$ \\ \hline
$2$ & & & & $Z$ & & & \\
$1$ & $\I\wh{q_0}$ & & $\I\wh{q_i}$ & & $\I\wh{p^i}$ & & $\I\wh{p^0}$ \\
$0$ & & $S^i$ & & $D^i_j, D, \Delta$ & & $T_i$ & \\
$-1$ & $\I\wh{Q_0}$ & & $\I\wh{Q_i}$ & & $\I\wh{P^i}$ & & $\I\wh{P^0}$ \\
$-2$ & & & & $\widetilde Z$ & & &
\end{tabular*}
\end{center}
\caption{A list of the generators of the Lie algebra $\fg'$, with two gradings, by $D$ (horizontal)
and by $\Delta$ (vertical).  This extends Figure \ref{fj-picture}, which was the analogous picture for $\tfg \subset \fg'$.} \label{g3-picture}
\end{figure}
Note in particular that $Z$ is not a central generator in $\fg'$:  together with $\widetilde Z$ and $\Delta$ it
generates an ${\mathfrak {sl}}(2,\R)$ subalgebra of $\fg'$.
The group $\tG$ is then the centralizer of $Z$ inside
$G'$.  This implies that if one begins with a representation $\cH'$ of $G'$ and then reduces $\cH'$ by
fixing $Z$ to some pure imaginary $\I \hbar$,\footnote{To be precise,
one defines $\cH_\hbar$ as a ``space of coinvariants,'' $\cH_\hbar = \cH / \{ Z \psi - \I \hbar \psi: \psi \in \cH \}$.}
the resulting space $\cH_\hbar$ will be a representation of $\tG$.

Having found the candidate $G'$, it remains to exhibit some $\cH'$ such that
the corresponding $\cH_\hbar$ is the \sw representation of $\tG$.
It turns out that the desired $\cH'$ is an analogue of the metaplectic representation of $Sp(4,\R)$, known
as the \ti{minimal} representation of $G'$.  These representations have been studied
extensively in the mathematics literature, \eg \cite{MR0342049,MR1159103,MR1195692,MR1278630,MR1327538,MR1372999,minspher,minrep-review},
and also by physicists, \eg \cite{Gunaydin:2000xr,Gunaydin:2005gd,Gunaydin:2005zz,Pioline:2001jn,Kazhdan:2001nx,Gunaydin:2001bt,Pioline:2004xq,Gunaydin:2004md,Pioline:2005vi,Gunaydin:2006vz}.
The action of the generators of $\fg'$ on functions in $n_v+2$ variables has been given explicitly
in \cite{Gunaydin:2005zz,Gunaydin:2004md,Gunaydin:2006vz} for simple groups
in their quaternionic real form, which is the case
presently relevant for us; however, the polarization chosen there was not the same as the
real polarization we used in describing the \sw representation above, making a direct comparison difficult.
Now we sketch a realization of the minimal representation which makes the relation more
transparent.  The representation space $\cH'$ consists of functions $f(p^I, y)$ in $n_v + 2$ variables.
The generators of $\tfg \subset \fg'$ are just rescaled versions of those in the \sw representation:
\begin{gather}
Z \mapsto \I y^2, \label{minrep-1} \\
\I\hq_0 \mapsto y \dwrt{p^0}, \quad \I\wh{q_i} \mapsto y \dwrt{p^i}, \quad \I\wh{p^i} \mapsto \I y p^i, \quad \I\wh{p^0} \mapsto \I y p^0,   \label{minrep-2} \\ \displaybreak[0]
S^i \mapsto -\ihalf C^{ijk} \frac{\pa^2}{\pa p^j \pa p^k} + p^i \dwrt{p^0}, \quad T_i \mapsto + \ihalf C_{ijk} p^j p^k - p^0 \dwrt{p^i}, \label{minrep-3} \\
R^j_i \mapsto \delta^j_i p^0 \dwrt{p^0} - p^j \dwrt{p^i} + \half C_{ikl} C^{jnl} \left(p^k \dwrt{p^n} + \dwrt{p^n} p^k\right). \label{minrep-4}
\end{gather}
In particular, fixing $y = 1$ we clearly recover the \sw representation of $\tG$ with $\hbar = 1$; more generally one could fix $y$ to any value, giving the
\sw representation with $\hbar = y^2$.
The other generators of $\fg'$ can also be described explicitly.
The generator $\Delta \in \fg'$ simply counts the degree in $y$,
\begin{equation}
\Delta \mapsto y \dwrt{y} + \half.
\end{equation}
To obtain the rest of $\fg'$ it is enough to give the action of $\widetilde Z$, since
$\I \wh{Q_I} =  [\widetilde Z, \I\wh{q_I}]$ and $\I \wh{P^I} = [\widetilde Z, \I\wh{p^I}]$;
this action can be given in terms of the quartic invariant,
\begin{equation}
\widetilde Z \mapsto \half \frac{\pa^2}{\pa y^2} - \frac{1}{4 y^2} \left( I_4(\wh{p^I}, \wh{q_I}) + \kappa \right),
\end{equation}
where $\kappa$ depends on the choice of ordering one makes in $I_4$.

Since we have seen that the topological string state $\ket{\Psi}$ lives in
the \sw representation $\cH_{\hbar = 1}$ of $\tG$, the existence of this extension of the \sw representation
suggests that there could also exist a one-parameter generalization of the topological string.  We will speculate
on its physical meaning in the next section.

So far we have been discussing the representation theory of the real groups $\tG$ and $G'$ without regard to their
integral subgroups.  But this discussion has an automorphic counterpart, which in our language amounts to
studying not just the relation between Hilbert spaces $\cH_\hbar$ and $\cH'$, but also some special states
which are invariant under discrete subgroups.  Such states give rise to automorphic forms for the corresponding groups.
The correspondence is easily described in the case
of $G' = Sp(4,\R)$:  given a holomorphic Siegel modular form $\Theta(t, \rho, z)$
for $Sp(4,\R)$ one can obtain holomorphic Jacobi forms for $SL(2,\R) \ltimes H$ by Fourier
expanding $\Theta$ in one of its three complex variables,
\begin{equation} \label{fj-expansion}
\Theta(t, \rho, z) = \sum_{m = 1}^\infty \theta_m(t, z) e^{\I m \rho}.
\end{equation}
Each $\theta_m(t, z)$ is then a Jacobi form of index $m$ \cite{MR781735}.
Now how about our case, where $Sp(4,\R)$ is replaced by $G'$?
The minimal representation of $G'$ is in some cases known to be automorphic, so there is a natural
candidate automorphic form $\Theta$, see \eg \cite{minspher,MR1767400,MR1469105}.
The automorphic extension of our discussion above would involve expanding $\Theta$ to
obtain Jacobi forms associated to the \sw representation of $\tG$.
As far as we know such an expansion is not in the mathematics literature, although some of
the necessary ingredients have been studied in \cite{marty-thesis}, which also discusses similar expansions for other ``small''
representations of $G'$.

\section{Discussion} \label{sec-discussion}

In this paper, we have shown that the holomorphic anomaly equations
for the topological string amplitude
can be rephrased as a heat equation for a purely holomorphic section,
\be
\label{heatsum}
\left[\dwrt{t^i} - \ihalf C_{ijk} \frac{\pa^2}{\pa y_{j} \pa y_{k}} + y_i \dwrt{w} \right]
\Psi_\hol(t; w, y^i) = 0.
\ee
Moreover, in the case where the moduli space $\cM$ is a
Hermitian symmetric tube domain $G/K$,
we have shown that the general solution can be written as a matrix element
\be
\Psi_\hol(t; y)
= \IP{\Psi | \exp (y_I \wh{p^I}) \exp \left(t^i T_i \right) | \Omega}
\ee
in the \sw representation of a non-reductive group $\tG = G\ltimes H$.
The holomorphic anomaly equation \eqref{heatsum} then follows from
identities in the annihilator of the \sw representation, which
moreover can be derived by embedding in the minimal representation
of a larger reductive group $G'$.

We close with a few comments and speculations.

\subsubsection*{An extended topological amplitude}

The realization of the holomorphic anomaly equation inside the \sw representation of $\tG$
is not surprising in light of the wave function interpretation.  In our opinion, though, its natural embedding
into $G'$ lends credence to the idea that there should exist a one-parameter
generalization of the usual topological
string amplitude, attached to the minimal representation of $G'$.
The extra parameter is $\hbar$, the eigenvalue of $Z$, which was fixed to $\hbar = 1$
in the \sw representation we used.

Now, what could the meaning of the extra parameter $\hbar$ be?  Here we present two speculations,
both relying on the fact that the group $G'$ can be interpreted physically
as the isometry group after
compactification on a circle to three dimensions.
First, note that by T-duality along the
circle (exchanging IIA and IIB) followed by decompactification to 4 dimensions,
$G'$ can be realized as the isometry group of a
hypermultiplet moduli space in 4 dimensions.
It is thus natural to conjecture that the ``generalized
topological string amplitude'' should compute the infinite series of
higher-derivative corrections $\tilde F_g$ on this hypermultiplet
moduli space.  The extra parameter $\hbar$ would then be related to the
presence of the ``universal hypermultiplet'' containing the physical string coupling,
controlling the strength of the spacetime instanton corrections.  To
make this proposal sharper it would be useful to have an off-shell superspace
description of the hypermultiplet derivative couplings, perhaps along the
lines of \cite{deWit:2006gn}.

For the other speculation we need to discuss how the
Heisenberg generators in $\tG$ are realized in $G'$.
They occur as shift symmetries for new scalars that come from the dimensional
reduction from $4$ to $3$ dimensions:  in particular, the
central generator $Z$ acts by shifting the scalar which arises from dualization of
the $d=3$ gauge field $g_{\mu t}$.  This scalar can be thought of as a dual potential
for the Taub-NUT charge, in which case fixing $Z$ is equivalent to fixing
the NUT charge of the 4-dimensional spacetime.  As was exploited in \cite{Dijkgraaf:2002ac,Kapustin:2004jm},
T-dualizing the space with $\hbar$ units of NUT charge leads to
$\hbar$ NS5-brane instantons wrapped on $X$
\cite{Ooguri:1996wj}.
In particular, suppose we start with the IIA theory; then we wind up with $\hbar$ NS5-brane instantons of
Type IIB.  In \cite{Kapustin:2004jm,Nekrasov:2004js} the $\hbar = 1$ case of this duality was used to argue that the
partition function of the single NS5-brane instanton captures the A model partition function on
$X$; this gave a physical interpretation to the results of \cite{Iqbal:2003ds,gw-dt,gw-dt2}, which argued
that the A model partition function can be computed using the maximally supersymmetric $U(1)$ gauge theory on $X$,
mathematically interpreted as the $U(1)$ ``Donaldson-Thomas'' theory \cite{thomas-thesis}.
This suggests that our ``generalized topological amplitude'' should include not only the $U(1)$ gauge theory
but also the higher $U(\hbar)$ theories for all $\hbar$; their
partition functions should be organized into
a single vector in the minimal representation of $G'$.

While these two proposals are both speculative, at least they
may be consistent with one another:  it is expected that the
hypermultiplet moduli space should be corrected by NS5-brane instantons,
so to say that the generalized
amplitude computes corrections to the hypermultiplet moduli space is not unrelated to saying it
computes the partition function of NS5-brane instantons. Moreover,
they are broadly consistent with the OSV conjecture, since the
instantons correcting the three-dimensional vector-multiplet moduli
space are four-dimensional black holes running along the thermal circle,
T-dual to the NS5-brane instantons.
We are hopeful that a proper understanding of the meaning of the
extended topological amplitude will lead to an exact form
of the OSV conjecture, valid at finite values of the charges,
along the lines of the ideas in
\cite{Ooguri:2005vr,Gunaydin:2005mx}.

\subsubsection*{Quantizing the intermediate Jacobian}

One of the main themes of this paper has been the analogy between the topological string partition function
$\Psi_\hol$ and the Jacobi theta series $\theta(\tau, z)$.  There is one crucial property
of $\theta(\tau, z)$ which we have mostly overlooked:
it transforms simply under shifts $z \to z + 1$, $z \to z + \tau$.
Geometrically, these transformations reflect the fact that
$\theta$ arises not from quantization of the linear space $H^1(T^2, \R)$
but from quantization of the Jacobian torus, $H^1(T^2, \R) / H^1(T^2, \Z)$.  Equivalently,
one can continue to think in terms of the linear space, but require invariance under
the integer form of the Heisenberg group.  This is a very strong constraint
on the state, determining it up to at most a finite-dimensional ambiguity.

From the point of view of the worldsheet of the topological string theory, there is no obvious reason
why the topological string state should have such an invariance.  On the other hand, if one takes seriously
the embedding $\tG \subset G'$ we advocated above, one finds that the phase space variables in the topological string get
a physical interpretation:  they are the Wilson lines of the $n_v+1$ gauge fields around the $d=4 \to d=3$ compactification
circle.
As such, they are naturally circle-valued, and one might expect that any partition function which depends on them
will have to be periodic in an appropriate sense.  One would then be led to consider quantization of the intermediate Jacobian
$J = H^3(X,\R) / H^3(X,\Z)$, with the corresponding theta functions as candidates for the full topological string
partition function.\footnote{This intermediate Jacobian has appeared in \cite{Witten:1997hc}, where it was argued that,
if one wraps an M-theory fivebrane on $X$, the dependence of its partition function on a flat background $C$ field
is given by a theta function similar to those we are discussing.
A similar argument was made in \cite{Dijkgraaf:2002ac} for the ``classical part'' of
the partition function of a stack of NS fivebranes in Type IIA.}

One
might well be skeptical of whether such a simple and universal object could really capture all the complexity of the topological
string theory on Calabi-Yau threefolds.  Moreover, even if the topological string state indeed turned out to be invariant under
shifts by $H^3(X,\Z)$, there would still be the problem of extracting the perturbative expansion.  This is not as trivial as it sounds:
as we discussed in Section \ref{sec-modular-symmetric}, in the complex structure we are using,
the geometric description of such a theta function would be as a class in sheaf cohomology $H^1$ rather than as a function.
Said differently, the series defining such a theta function is badly divergent.
Still, the case of a Calabi-Yau one-fold gives some ground for optimism:
the partition function which counts covering maps to the torus can
indeed be expressed in terms of
a ``generalized Jacobi form,'' defined by a formal product which
definite modular properties \cite{MR1363055,MR1363056};
in particular, the
$SL(2,\Z)$ transformation of the variables seems to have a natural
interpretation in terms of quantization of a phase space in the 4-dimensional
representation of $SL(2,\R)$.

\subsubsection*{Extensions to $\N=4$ and $\N=8$}

In this paper we have been concerned only with the case of $\N=2$ supersymmetry.  Given
our focus on cases where the moduli space is a symmetric space, it might have seemed more natural to consider
the $\N=4$, $\N=6$ and $\N=8$ theories first, since in those cases the moduli space is always a symmetric space,
which we again write as $G' / K'$ in $d=3$.  Then one can
construct the minimal representation of $G'$; its matrix elements are very special functions on $G'$,
obeying a large number of differential equations, which could be easily written down
much as we did above in the $\N=2$ case.  It is natural to wonder whether these equations
arise somewhere in physics.

There is an important difference between the $\N=2$ supergravities defined by cubic Jordan algebras,
which we considered above, and the $\N=8$ theory or the $\N=4$ theories coupled to vector multiplets.
Namely, in general the groups $G$ occurring in the latter cases in $d=4$ do not have holomorphic discrete series
representations.  Roughly this means one should not expect to construct holomorphic ``wave functions''
from the representations of $G$.
So the analogues of the anomaly equations which appear must have a somewhat
subtler interpretation.  One notable exception is the $\N=4$ supergravity coupled to $2$ vector multiplets, where the moduli space in $d=4$ is $(SL(2,\R) \times SO(2,6)) / (U(1) \times U(1) \times SO(6))$.  The existence of holomorphic discrete series in this case may be related to the fact that this theory can be obtained by consistent truncation
from the $\N=6$ theory, for which the corresponding group $SO^*(12)$ does admit holomorphic discrete series
representations.

In the case of Type II string theory compactified on K3 one might ask whether the
differential equations one obtains are related to the
holomorphic anomaly of the $\N=4$ topological string on K3 \cite{Berkovits:1995vy,Ooguri:1995cp}.
To obtain such a relation one would presumably
have to decompactify from $d=3$ to $d=6$, which should correspond to fixing $3$ generators of $G'$
(analogous to the $\N=2$ case where we decompactified from $d=3$ to $d=4$ and fixed the single generator $Z$.)
The centralizer of these generators in $G' = O(24,8)$ should be some nilpotent extension $\tG = O(20,4) \ltimes N$,
where $O(20,4)$ arises as the isometry group of the moduli space in $d=6$.
Then the natural conjecture, analogous to what we found in the $\N=2$ case,
is that the anomaly equations of the $\N=4$ topological string
can be interpreted as statements about some representation of this $\tG$.

One can similarly ask whether any meaning can be attached to the differential equations which arise
in the $\N=8$ case from the minimal representation of $E_{8(8)}$.

\section*{Acknowledgements}
We are grateful to M.~Aganagic, R.~Dijkgraaf, R.~Donagi, T.~Grimm, B.~Gross, M.~Headrick, S.~Hellerman,
A.~Klemm, G.~Moore, L.~Silberman, E.~Verlinde, M.~Weissman, A.~Wienhard and E.~Witten
for useful discussions.
A.~N. and B.~P. thank CIRM for hospitality during the final part of this
work. The research of
A.~N. is supported by the Martin A. and Helen Chooljian Membership at the
Institute for Advanced Study and by NSF grant PHY-0503584.
The research of B.P. is supported by the EU under contracts
MTRN--CT--2004--005104, MTRN--CT--2004--512194, and by
ANR (CNRS--USAR) contract No 05--BLAN--0079--01.  The work of M.~G. was supported
in part by the National Science Foundation under grant number PHY-0555605.  Any
opinions, findings and conclusions or recommendations expressed in this material
are those of the authors and do not necessarily reflect the views of the National
Science Foundation.

\appendix

\section{Holomorphic ambiguities and real polarization}

In this Appendix, we discuss the relation between the ``holomorphic
ambiguities'' found by BCOV in their procedure for solving
the holomorphic anomaly equations recursively and the real
polarized topological amplitude defined in \eqref{psipwy}.

In Section 6.2 of \cite{Bershadsky:1994cx}, BCOV
introduce a variant of \eqref{psibcov},
\be
\label{psibcovt}
\tilde\Psi_{\rm BCOV} (t^i,\bar t^{\bar j}; \tilde x^i, \varphi; \tilde\lambda)
= \tilde\lambda^{1-\frac{\chi}{24}}\
\exp\left[ W\left(t^i,\bar t^{\bar j}; \frac{\tilde x^i}{1-\varphi},
\frac{\tilde\lambda}{1-\varphi} \right) \right].
\ee
Here $\tilde\lambda$ and $\varphi$ are
redundant variables: $\varphi$ describes variations of the
inverse topological coupling $\lambda^{-1}$, while
$\tilde\lambda$ only serves to
organize the perturbative series.
Then the first anomaly equation \eqref{BCOV1} becomes
\be
\label{BCOV1t}
\pa_{\bar t^i} =
\frac{\tilde\lambda^2}{2} e^{2K} \bar C_{\bar i\bar j\bar k}
g^{j\bar j} g^{k\bar k}
\frac{\pa^2}{\pa \tilde x^j\pa \tilde x^k}
- g_{\bar i j} \tilde x^j \frac{\pa}{\pa \varphi}
\ee
Moreover, BCOV observe that the ``conjugate'' equation
\be
\label{BCOV1tc}
\pa_{\bar t^i} =
- \frac{\tilde\lambda^2}{2} e^{2K} \bar C_{\bar i\bar j\bar k}
g^{j\bar j} g^{k\bar k}
\frac{\pa^2}{\pa \tilde x^j\pa \tilde x^k}
- g_{\bar i j} \tilde x^j \frac{\pa}{\pa \varphi}
\ee
is solved by
\be
\label{psicl}
\tilde\Psi_{Y}
(t,\bar t; \tilde\lambda, \tilde x, \varphi)
= \exp\left[ -\frac{1}{2\tilde\lambda^2} \left(
\Delta_{ij} \tilde x^i \tilde x^j
+ 2 \Delta_{i\varphi} \tilde x^i \varphi + \Delta_{\varphi\varphi} \varphi^2
\right)
+ \half \log\left(\frac{\det\Delta}{\tilde\lambda^2} \right) \right]
\ee
where the ``inverse propagator'' $\Delta$ is defined to obey the equations
\be
\label{invprop}
\Delta= \begin{pmatrix}
\Delta_{\varphi\varphi} & \Delta_{i\varphi} \\
\Delta_{i\varphi} & \Delta_{ij}
\end{pmatrix}
,\quad
\Delta^{-1} =
\begin{pmatrix}
-2 S  & - S^i \\
-S^i & - S^{ij}
\end{pmatrix}
=
\begin{pmatrix}
-2 S  & - g^{i\bar j} \pa_{\bar j} S \\
- g^{i\bar j} \pa_{\bar j} S &
- g^{i\bar j} \pa_{\bar j} ( g^{j\bar k} \pa_{\bar k} S )
\end{pmatrix},
\ee
and $S$ is a local section of ${\cal L}^{-2}$ whose third derivative
reproduces the tree-level three-point function,
\be
\label{cds}
C_{\bar i\bar j\bar k} = e^{-2K} D_{\bar i} D_{\bar j} D_{\bar k} S.
\ee
Note that this does not define $S$ unambiguously.
Using \eqref{BCOV1t}, \eqref{BCOV1tc} BCOV conclude that the integral
\be
\label{holam}
Z(t,\bar t) = \int d\tilde x d\varphi ~
\tilde\Psi_{\rm BCOV}(t,\bar t; \tilde x, \tilde \lambda, \varphi)~
\Psi_{Y}(t,\bar t; \tilde x, \tilde \lambda, \varphi)
\ee
is in fact independent of $\bar t$, and
identify it as the generating function of the ``holomorphic
ambiguities'' which arise at each genus upon integrating the
first anomaly equation.

Now let us relate
this construction to the intertwiner from Verlinde's $\psiv$ to the
real polarized $\Psi_\R$.
The first thing to notice is that Eq.~\eqref{BCOV1t} is identical
to \eqref{Ver1b} after shifting $\varphi\to \varphi+1$ in \eqref{psibcovt},
setting $\tilde\lambda=1$ and identifying $\varphi= - \lambda^{-1}$.
Next, we invert the formula \eqref{psixpsip} which gave the Bogoliubov
transformation from $\Psi_\R$ to $\psiv$, obtaining
\be
\label{psippsix}
\Psi_{\IR}(p^I)=  \int dx^I~d\bar x^I~\sqrt{\det[\im\tau]}
~e^{-\half x^I [\im\tau]_{IJ} \bar x^J} ~
\psiv^*(p^I; X^I,\bar X^I; \bar x^I)~\psiv(X^I,\bar X^I; x^I).
\ee
The integral over $\bar x^I$ is easily evaluated (at least formally),
leading to
\be
\label{psipwy}
\begin{split}
\Psi_{\IR}(p^I)=&  \int dx^I~\sqrt{\det[\im\tau]}
\exp\left[ \ihalf p^I \bar\tau_{IJ} p^J
- p^I [\im\tau]_{IJ} x^J
+ \qtr x^I [\im\tau]_{IJ} x^J  \right]\\
& \times ~\psiv(X^I,\bar X^I; x^I).
\end{split}
\ee
Setting $p^I=0$ in this expression, and replacing the large phase space
variables $x^I$ in terms of $(x^i,\lambda)$ in the exponent, we find
\begin{multline}
\label{twinerdvv}
\Psi_{\IR}(p^I)=  \int d\lambda^{-1}dx^i~ \psiv(X^I,\bar X^I; x^i,\lambda)~ \sqrt{\det[\im\tau]}\\
\exp\left[
\frac14  \left( \lambda^{-2} X^I [\im\tau]_{IJ} X^J
+2 \lambda^{-1} e^{-K/2} x^i f_i^I [\im\tau]_{IJ} X^J
+ e^{-K} x^i f_i^I [\im\tau]_{IJ} f_j^J x^j \right) \right].
\end{multline}
This coincides with \eqref{psicl} if we identify
\bea
\label{Dgen}
\Delta_{ij} &=& -2 e^{-K} f_i^I [\im\tau]_{IJ}  f_j^J,  \\
\Delta_{i\varphi} &=& 2 e^{-K/2} f_i^I [\im\tau]_{IJ} X^J,  \\
\Delta_{\varphi\varphi} &=& -2 X^I [\im\tau]_{IJ} X^J,
\eea
where we introduced the standard special geometry notation $f_i^I := e^{K/2} D_i X^I$.

Inverting the matrix $\Delta$ with these entries, we obtain
\bea
\label{Sgen}
S &=& e^{2K}  \bar X^I [\im\tau]_{IJ} \bar X^J, \\
S_{\bar i}&=& 2 e^{3K/2} ~ \bar f^I_{\bar i} [\im\tau]_{IJ} \bar X^J, \\
S_{\bar i\bar j} &=& 2 e^{K}
\bar f^I_{\bar i} [\im\tau]_{IJ} \bar f^J_{\bar j}.
\eea
Then using the special geometry identities
\be
X^I (\pa_i [\im\tau]_{IJ}) X^J =
f^I_i (\pa_j [\im\tau]_{IJ}) X^J = 0,\quad
f^I_i (\pa_j [\im\tau]_{IJ}) f_k^J = -\ihalf e^{-K} C_{ijk},
\ee
we check that $S$ given by \eqref{Sgen} indeed satisfies \eqref{cds},
and moreover that $S_{\bar i}=D_{\bar i} S, S_{\bar i\bar j}=
D_{\bar i} D_{\bar j}S$ as required in \eqref{invprop}.

We conclude that, for the $S$ given in \eqref{Sgen}, $Z(t)$ is proportional
to the wave function in the real polarization at $p^I=0$, namely
\be
\label{zf1}
Z(t) = e^{f_1(t)}  \Psi_{\IR}(0)
\ee
where the exponential factor arises from the redefinition \eqref{psiv}.
We emphasize that \eqref{zf1} holds only for the
$S$ given in \eqref{Sgen}, which need not be globally
well defined; for the choice of $S$ made in \cite{Bershadsky:1994cx},
$Z(t)$ in fact contains the infinite series of holomorphic ambiguities.

\section{Wave function in the positive-definite polarization}

With a little more work, we can also determine the intertwiner $\psiw$ from the
real polarization to a
positive definite (``Weil'') polarization given by the Hodge $\star$:
as discussed below \eqref{weilj},
this $\psiw$ should be obtained from the intertwiner $\psiv$ to the
indefinite (``Griffiths'') polarization by Fourier transforming from
$\lambda^{-1}$ to a new variable $\mu$.
So we begin from the formula \eqref{intert} for $\psiv$, replace the large phase space variables $x^I$ with
the small phase space ones $(x^i, \lambda)$ using \eqref{lphase}, and then formally Fourier transform
\begin{equation}
\psiw(p^I; X^I, \bar{X}^I; x^i, \mu) = \int d\lambda^{-1} \exp \left( \lambda^{-1} \mu \right) \psiv(p^I; X^I, \bar{X}^I; x^i, \lambda).
\end{equation}
We thus obtain $\psiw$ in terms of the negative-definite period matrix
\be
{\cal N}_{IJ} = \bar\tau_{IJ} + 2\I \frac{ [\im\tau]_{IK} X^K ~
[\im\tau]_{JL} X^L}{X^K[\im\tau]_{KL}X^L},
\ee
namely, we find
\be
\label{twinerdvvf}
\begin{split}
\psiw(p^I;X^I,\bar X^I; x^i,\mu) =&
\frac{\sqrt{\det[\im\tau]}}{\sqrt{X^I [\im\tau]_{IJ} X^J}}
\exp\left[
\ihalf p^I {\cal N}_{IJ} p^J
- 4 {\rm i} e^{K} p^I [\im\N]_{IJ} \bar X^J \mu \right.\\
&\left.
-e^{-K/2} p^I [\im\N]_{IJ} f^J_j x^j
+ 4 e^{2K} \bar X^I [\im\N]_{IJ} \bar X^J \mu^2 \right. \\
& \left.
- 2 i e^{K/2}~f_i^I [\im\N]_{IJ} \bar X^J  x^i \mu
- \frac14 e^{-K} f_i^I [\im\N]_{IJ} f_j^J x^i x^j \right].
\end{split}
\ee
Here we used the special geometry properties
\bea
\bar X^I [\im\N]_{IJ} \bar X^J &=& \frac14
\frac{e^{-2K}}{X^I [\im\tau]_{IJ} X^J} \\
f^I_i [\im\N]_{IJ} \bar X^J &=& -\frac12
\frac{e^{-K}\ f^I_i [\im\tau]_{IJ} X^J}{X^I [\im\tau]_{IJ} X^J} \\
f^I_i [\im\N]_{IJ} f^J_j &=&
\frac{ (f^I_i [\im\N]_{IK} X^K) (f^J_j [\im\N]_{JL} X^L)}
{X^I [\im\tau]_{IJ} X^J} - f^I_i [\im\tau]_{IJ} f_j^J \\
p^I [\im\N]_{IJ} \bar X^J &=& -\frac12 e^{-K}
\frac{p^I [\im\tau]_{IJ}X^J}{X^I [\im\tau]_{IJ} X^J}
\eea
Defining new large phase space coordinates (for $H^{1,2} \oplus H^{0,3}$) by
\be
w^I = 2 \I\  \mu \bar X^I e^K + \frac12 x^i D_i X^I,
\ee
the intertwiner between the real and positive-definite
polarizations takes the simple form
\be
\label{twinerdvvff}
\psiw(p^I;X^I,\bar X^I; w^I) =
\frac{\sqrt{\det[\im\tau]}}{\sqrt{X^I [\im\tau]_{IJ} X^J}}
\exp\left[
\ihalf p^I {\cal N}_{IJ} p^J  -2 p^I [\im\N]_{IJ} w^J
- w^I  [\im\N]_{IJ} w^J
\right]
\ee
which is closely similar to \eqref{intert}.

\renewcommand{\baselinestretch}{1}
\small\normalsize

\bibliography{combined}

\end{document}